\documentclass[prc,twocolumn]{revtex4-1}

\usepackage{enumerate}

\usepackage{booktabs}
\usepackage{color}
\usepackage{graphicx}

\usepackage{amsmath}
\usepackage{amssymb}
\usepackage{times}
\usepackage{multirow}
\usepackage{makecell}
\def\bvec#1{\mbox{\boldmath $#1$}}

\newcommand{\del}[2]{\frac{\partial #1}{\partial #2}}

\newcommand{\bra}{\langle}
\newcommand{\ket}{\rangle}

\usepackage{ascmac}
\usepackage{fancybox}

\newcommand{\beq}{\begin{equation}}
\newcommand{\eeq}{\end{equation}}
\newcommand{\bea}{\begin{eqnarray}}
\newcommand{\eea}{\end{eqnarray}}

\def\fun#1#2{\lower3.6pt\vbox{\baselineskip0pt\lineskip.9pt
 \ialign{$\mathsurround=0pt#1\hfil##\hfil$\crcr#2\crcr\sim\crcr}}}

\begin{document}

\title{
  Microscopic Characterization of Nuclear Collectivity and Branching Ratios \\
  from Jost-RPA Pole Residues
}

\author{Kazuhito Mizuyama$^{1,2}$}
\email{corresponding author: mizukazu147@gmail.com}

\affiliation{
  \textsuperscript{1}
  Institute of Research and Development, Duy Tan University,
  Da Nang 550000, Vietnam
  \\
  \textsuperscript{2}
  Faculty of Natural Sciences,  Duy Tan University, Da Nang 550000, Vietnam
}

\date{\today}

\begin{abstract}
Within the continuum Random Phase Approximation formulated via the Jost function method (Jost-RPA), we formulate a microscopic method to (i) evaluate nuclear collectivity through a dimensionless metric normalized to unity for a pure single-channel excitation, and (ii) extract partial decay widths and branching ratios directly from complex channel amplitudes without artificial width broadening. 
By applying rank-1 Takagi factorization to the complex symmetric scattering matrix residue at resonance poles across multiple Riemann sheets, the transition strength residue is algebraically partitioned into the single-particle radial overlap scale, multi-channel participation coherence, and operator phase alignment. 
To validate the effectiveness and physical consistency of this framework, benchmark calculations are performed for $^{16}\text{O}$ across diverse continuum excitation modes: 
(i) for the isovector giant dipole resonance, the metric identifies a strongly multi-channel collective response decaying concurrently through multiple open channels; 
(ii) for the isoscalar quadrupole channel, the metric characterizes the strong collectivity of the low-lying resonance concentrating the low-energy transition strength, while higher core-excited states exhibit suppressed direct excitation and destructive continuum interference; and 
(iii) for the isovector quadrupole channel, the metric clearly reveals that the most prominent strength peak is a non-collective quasi-single-particle resonance trapped behind a centrifugal barrier, whereas collective IVGQR components are pushed up and fragmented across higher-energy Riemann sheets. 
This work establishes a transparent diagnostic tool for characterizing continuum resonances and decay branching, providing a practical framework for future applications to heavier nuclei where collective phenomena fully develop.
\end{abstract}

\maketitle

\section{Introduction}
\label{sec:intro}

Nuclear giant resonances, such as the isovector giant dipole resonance (GDR) and the isoscalar and isovector giant quadrupole resonances (ISGQR and IVGQR), represent fundamental collective excitation modes of atomic nuclei across the periodic table \cite{Speth1981}. 
Microscopically, these giant modes are understood as coherent superpositions of numerous particle-hole ($1p$-$1h$) excitations driven by residual nucleon-nucleon interactions. 
In bound-state shell-model and discretized Random Phase Approximation (RPA) frameworks, nuclear collectivity has historically been evaluated by the enhancement of reduced transition probabilities $B(EL)$ relative to single-particle Weisskopf estimates, under the premise that an unperturbed single-particle transition cannot exceed the single-particle limit. 
In parallel, statistical and information-theoretic measures, such as the participation ratio (PR) and information entropy \cite{Wegner1980, Armstrong2012, Borgonovi2016}, have widely been employed to quantify configuration mixing and wave-function complexity.

However, because giant resonances typically lie well above particle emission thresholds, their consistent microscopic description requires an open-quantum-system treatment that explicitly accounts for continuum boundary conditions and particle decay escape widths ($\Gamma^\uparrow$). 
Standard discrete RPA approaches using harmonic oscillator bases or artificial box boundary conditions discretize the continuum into discrete pseudostates, relying on empirical Lorentzian width parameters ($\Delta$) to generate smooth strength functions. While this procedure simulates the overall strength distribution, it obscures individual complex resonance poles ($E_g = \operatorname{Re}E_g - i\Gamma_g/2$ with $\Gamma_g>0$), escape widths, and partial decay branching ratios into specific residual channels.

Furthermore, analyzing nuclear collectivity in the open continuum presents fundamental challenges that make traditional criteria ambiguous:
\begin{enumerate}[(i)]
\item \emph{Spatial overlap enhancement versus multi-channel collectivity:} Unperturbed single-particle transitions into continuum waves with high orbital angular momentum can produce large transition strengths solely through substantial spatial wave-function overlaps trapped behind centrifugal barriers, generating prominent peaks without multi-channel configuration mixing.
\item \emph{Phase coherence versus incoherent fragmentation:} Probability-based measures like the participation ratio depend only on squared configuration amplitudes, discarding relative phase information. Consequently, they cannot distinguish in-phase constructive interference (coherent collective vibrations) from out-of-phase or random cancellations (incoherent continuum spreading and fragmentation).
\end{enumerate}

To overcome the limitations of discretized continuum models, coordinate-space continuum RPA approaches formulated via the Jost function method (JRPA) \cite{Mizuyama2023_JRPA1, Mizuyama2024_Eigenphase, Mizuyama2025_JRPA2} have recently been developed. The Jost RPA constructs the coordinate-space outgoing-wave Green's function and complex symmetric scattering matrix $\mathcal{S}(E)$ directly, providing direct access to complex resonance poles $E_g$ across multiple Riemann sheets without box boundary conditions or artificial width broadening.

Rather than seeking a universal philosophical definition of nuclear collectivity, the primary objective of this work is to establish a practical microscopic diagnostic tool tailored to open-channel continuum RPA calculations. 
By utilizing the complex symmetry of the S-matrix residue at isolated simple resonance poles, we apply rank-1 Takagi factorization \cite{Autonne1915, Takagi1924, Mizuyama2026_Takagi, Guo2016} to extract complex channel amplitudes that consistently preserve both partial decay widths and relative branching ratios along with channel phases. 
We algebraically partition the transition strength residue into the single-particle radial overlap scale, multi-channel participation coherence $\eta_g$, and operator phase alignment $C_g^2$, constructing a dimensionless collectivity metric $C_g^2 N_{\text{tot}}\eta_g$ normalized to unity at the single-particle limit.

We apply this framework to the doubly magic nucleus $^{16}\text{O}$ as a practical benchmark within the $1p$-$1h$ continuum RPA using a phenomenological Woods-Saxon mean field and Skyrme residual interactions. 
This application serves two main purposes: first, the manageable channel dimension of $^{16}\text{O}$ allows a transparent channel-by-channel verification of the complex amplitudes across diverse continuum excitation modes (isovector dipole, isoscalar quadrupole, and isovector quadrupole); second, it provides an established baseline from earlier JRPA calculations to validate our modernized computational implementation—which incorporates the multichannel Renormalized Numerov method \cite{Johnson1977} and the contour integral root-finding (CIRR) method \cite{Sakurai2003} developed to prepare for future large-scale applications to heavy nuclei.

The paper is organized as follows: Sec.~\ref{sec:formalism} formulates the rank-1 Takagi decomposition and collectivity metrics. Sec.~\ref{sec:results} presents numerical results and microscopic decay analyses for $^{16}\text{O}$. Sec.~\ref{sec:summary} summarizes the findings. Appendices~\ref{app:jost_formalism} and \ref{app:jost_green} outline the underlying Jost RPA equations and Green's functions.

\section{Theoretical formalism}
\label{sec:formalism}

The microscopic description of continuum excitations within the Random Phase Approximation (RPA) requires a proper treatment of boundary conditions in open quantum systems. In this work, the underlying coupled-channel radial differential equations, Woods-Saxon mean-field potentials, Skyrme-type residual particle-hole interactions, and the complete construction of the coordinate-space Jost-RPA Green's function $\mathcal{G}(r,r')$ and the symmetric flux-adjusted S-matrix $\mathcal{S}(E)$ \cite{Mizuyama2023_JRPA1, Mizuyama2024_Eigenphase, Mizuyama2025_JRPA2} are summarized in Appendices~\ref{app:jost_formalism} and \ref{app:jost_green}. In this section, we focus on the extraction of resonance pole properties via the Takagi factorization of the S-matrix residue, and formulate the collectivity indices, channel participation rates, and relative branching ratios derived from the complex pole amplitudes.

\subsection{Takagi factorization and S-matrix residue}
\label{subsec:takagi_residue}
Since the flux-adjusted S-matrix defined in Eq.~(\ref{eq:Smat-F}) of Appendix~\ref{app:jost_green} is a complex symmetric matrix,
it can be decomposed as
\begin{equation}
  \mathcal{S}
  =
  W\Sigma W^{\mathsf{T}}
\end{equation}
using the Takagi decomposition \cite{Autonne1915, Takagi1924}, where $W$ is a $2N_{\text{tot}}\times 2N_{\text{tot}}$ unitary matrix
and $\Sigma$ is a real non-negative diagonal matrix:
$\Sigma=\operatorname{diag}(\sigma_1,\dots,\sigma_{2N_{\text{tot}}})$.
Since the residue at a pole of the S-matrix is a complex symmetric matrix of rank-1,
and there exists only one nonzero, non-negative singular value in the Takagi decomposition of
the residue at a pole of the S-matrix, the Takagi decomposition of the residue at a pole of
the S-matrix is given as
\begin{equation}
  \operatorname{Res}\left[\mathcal{S},E_g\right]
  =
  \sigma_g
  \vec{\omega}_g\vec{\omega}_g^{\mathsf{T}}
  \label{takagi1}
\end{equation}
where $\vec{\omega}_g$ is a $2N_{\text{tot}}$-dimensional Takagi vector normalized as $\|\vec{\omega}_g\|=1$. In this framework, the resonance poles are assumed to be isolated and simple (non-degenerate). Numerically, the largest singular value $\sigma_g^{(c)}$ is retained, while higher singular values ($\sigma_2, \sigma_3, \dots$) arising from numerical discretization are verified to be negligible. Indeed, for all resonance poles investigated in the present work, the ratio of the second singular value to the dominant first singular value satisfies $\sigma_2 / \sigma_1 \le 10^{-9}$ (and in most cases $\sim 10^{-13}$), confirming that the S-matrix residue at each resonance pole is strictly rank-1 down to the limits of numerical precision.

Under the condition of an isolated simple pole (where the S-matrix residue has rank 1), the Takagi vector $\vec{\omega}_g$ is equivalent to the asymptotic partial width amplitudes of the resonant Gamow states with biorthogonal normalization \cite{Mizuyama2026_Takagi}. Furthermore, extracting resonance coupling strengths directly from the residues of scattering S-matrix poles provides a non-perturbative characterization of continuum resonances \cite{Guo2016}.

In open quantum systems with both open continuum channels and closed/decay channels, the asymptotic amplitudes of the unphysical closed channels diverge exponentially in coordinate space. Consequently, although the full scattering matrix $\mathcal{S}(E)$ is formally symmetric, the matrix elements in the closed-channel sub-block $\mathcal{S}_{dd}$ exceed those of the open-channel block $\mathcal{S}_{cc}$ by many orders of magnitude. Direct numerical application of the Takagi factorization to the entire $2N_{\text{tot}} \times 2N_{\text{tot}}$ matrix is therefore hindered by severe numerical instabilities and precision overflow. To overcome this limitation, one can exploit the coupled-channel boundary equations to establish an exact algebraic reduction that projects the global singular value and vector from the stable open-channel sub-block $\mathcal{S}_{cc}$.

Let the Takagi singular value and normalized singular vector associated with the open-channel residue $\operatorname{Res}\left[\mathcal{S}_{cc},E_g\right]$ be denoted by $\sigma_g^{(c)}$ and $\vec{\omega}_g^{(c)}$ ($\|\vec{\omega}_g^{(c)}\|=1$), such that
\begin{equation}
  \operatorname{Res}\left[\mathcal{S}_{cc},E_g\right]
  =
  \sigma_g^{(c)}
  \vec{\omega}_g^{(c)}
  \vec{\omega}_g^{(c)\mathsf{T}}.
  \label{takagi2}
\end{equation}
The residues of the other blocks of the S-matrix can be expressed as
\begin{align}
  \operatorname{Res}\left[\mathcal{S}_{cd},E_g\right]
  &=
  \operatorname{Res}\left[\mathcal{S}_{cc},E_g\right]
  \left(\mathcal{M}_{cc}^{(-)}\right)^{-1}
  \mathcal{N}_{cd}
  \\
  \operatorname{Res}\left[\mathcal{S}_{dc},E_g\right]
  &=
  -
  \left(\mathcal{J}_{dd}^{(+)}\right)^{-1}
  \mathcal{J}_{dc}^{(+)}
  \operatorname{Res}\left[\mathcal{S}_{cc},E_g\right]
  \\
  \operatorname{Res}\left[\mathcal{S}_{dd},E_g\right]
  &=
  -
  \left(\mathcal{J}_{dd}^{(+)}\right)^{-1}
  \mathcal{J}_{dc}^{(+)}
  \operatorname{Res}\left[\mathcal{S}_{cc},E_g\right]
  \nonumber\\
  &\times
  \left(\mathcal{M}_{cc}^{(-)}\right)^{-1}
  \mathcal{N}_{cd}
\end{align}
where
\begin{align}
  \mathcal{M}_{cc}^{(\pm)}
  &\equiv
  \mathcal{K}_c^{\frac{1}{2}}
  \left[
    \mathcal{J}_{cc}^{(\pm)}
    -
    \mathcal{J}_{cd}^{(+)}
    \left(\mathcal{J}_{dd}^{(+)}\right)^{-1}
    \mathcal{J}_{dc}^{(\pm)}
    \right]
  \mathcal{K}_c^{-\frac{1}{2}}
  \\
  \mathcal{N}_{cd}
  &\equiv
  \mathcal{K}_c^{\frac{1}{2}}
  \left[
    \mathcal{J}_{cd}^{(-)}
    -
    \mathcal{J}_{cd}^{(+)}
    \left(\mathcal{J}_{dd}^{(+)}\right)^{-1}
    \mathcal{J}_{dd}^{(-)}
    \right]
  \mathcal{K}_d^{-\frac{1}{2}}
\end{align}
where the Jost matrices and channel momenta are partitioned according to open ($c$) and closed ($d$) channel subspaces as
\begin{equation}
  \mathcal{J}^{(\pm)}
  =
  \begin{pmatrix}
    \mathcal{J}_{cc}^{(\pm)} & \mathcal{J}_{cd}^{(\pm)} \\
    \mathcal{J}_{dc}^{(\pm)} & \mathcal{J}_{dd}^{(\pm)} 
  \end{pmatrix},
  \hspace{5pt}
  \mathcal{K}
  =
  \begin{pmatrix}
    \mathcal{K}_{c} & 0 \\
    0& \mathcal{K}_{d} 
  \end{pmatrix}.
\end{equation}
From the intrinsic symmetry of the scattering matrix $\mathcal{S}^{\mathsf{T}} = \mathcal{S}$ and its pole residue, the off-diagonal sub-blocks satisfy the exact relation
\begin{equation}
  \mathcal{N}_{cd}^{\mathsf{T}}
  \left(\mathcal{M}_{cc}^{(-)-1}\right)^{\mathsf{T}}
  =
  -
  \left(\mathcal{J}_{dd}^{(+)}\right)^{-1}
  \mathcal{J}_{dc}^{(+)}.
\end{equation}
Substituting this algebraic relation into the residue block expansion directly establishes the rank-1 factorization for the entire $2N_{\text{tot}} \times 2N_{\text{tot}}$ residue matrix:
\begin{equation}
  \operatorname{Res}\left[\mathcal{S},E_g\right]
  =
  \sigma_g^{(c)}
  \vec{q}_g
  \vec{q}_g^{\mathsf{T}}
  \label{takagi3}
\end{equation}
where
\begin{equation}
  \vec{q}_g
  \equiv
  \begin{pmatrix}
    \vec{\omega}_g^{(c)}
    \\
    \mathcal{N}_{cd}^{\mathsf{T}}
    \left(\mathcal{M}_{cc}^{(-)-1}\right)^{\mathsf{T}}
    \vec{\omega}_g^{(c)}
  \end{pmatrix}.
\end{equation}
Although the composite vector $\vec{q}_g$ spans the full $2N_{\text{tot}}$-dimensional particle-hole space, it is not normalized ($\|\vec{q}_g\| \neq 1$). By matching Eq.~(\ref{takagi3}) with the canonical normalized Takagi factorization in Eq.~(\ref{takagi1}), the physical normalized Takagi vector $\vec{\omega}_g$ and the total singular value $\sigma_g$ are uniquely obtained as
\begin{equation}
  \vec{\omega}_g
  =
  \frac{\vec{q}_g}{\|\vec{q}_g\|},
  \hspace{5pt}
  \sigma_g
  =
  \sigma_g^{(c)}\|\vec{q}_g\|^2.
\end{equation}

\subsection{Residue of the RPA response function}

By inserting the open-channel Takagi factorization of the S-matrix residue [Eq.~(\ref{takagi2})] into the coordinate-space representation of the Jost-RPA Green's function [Eq.~(\ref{Green-F2}) of Appendix~\ref{app:jost_green}], the residue of the Green's function at the complex resonance pole $E_g$ factorizes into a dyadic form:
\begin{align}
  &
  \operatorname{Res}\left[\mathcal{G}(r,r'),E_g\right]
  \nonumber\\
  &=
  -i\frac{4m\sigma_g^{(c)}}{\hbar^2}
  \Psi_{F,c}^{(+)*}(r;E^*_g)
  \vec{\omega}_g^{(c)}
  \vec{\omega}_g^{(c)\mathsf{T}}
  \Psi_{F,c}^{(+)\dagger}(r';E^*_g)
  \nonumber\\
  &=
  \pi
  \vec{\psi}_g(r)
  \vec{\psi}_g^{\mathsf{T}}(r')
\end{align}
where $\vec{\psi}_g(r)$ is a $2N_{\text{tot}}$-dimensional vector defined by
\begin{equation}
  \vec{\psi}_g(r)
  \equiv
  \sqrt{\frac{1}{\pi i}\frac{4m\sigma_g^{(c)}}{\hbar^2}}
  \Psi_{F,c}^{(+)*}(r;E^*_g)
  \vec{\omega}_g^{(c)}
\end{equation}
The radial vector field $\vec{\psi}_g(r)$ defines the coordinate-space resonant state wave function associated with the pole $E_g$. It naturally splits into forward ($X$) and backward ($Y$) particle-hole components of dimension $N_{\text{tot}}$:
\begin{equation}
  \vec{\psi}_g(r)
  =
  \begin{pmatrix}
    \vec{\psi}_{g,X}(r) \\
    \vec{\psi}_{g,Y}(r)
  \end{pmatrix}.
\end{equation}
The neutron and proton transition densities $\rho_{q}^{(g)}(r)$ ($q=n,p$) at the resonance pole $E_g$ are constructed by contracting $\vec{\psi}_g(r)$ with the corresponding single-particle hole basis vectors:
\begin{align}
  \rho_{n}^{(g)}(r)
  &\equiv
  \vec{\psi}_g^{\mathsf{T}}(r)
  \xi_{n}(r)
  \label{trddef1n}
  \\
  &=
  \sum_{i=1}^{N_n}
  \left(
  \vec{\psi}_{g,X}^{\mathsf{T}}(r)
  +
  \vec{\psi}_{g,Y}^{\mathsf{T}}(r)
  \right)_i
  \tilde{\varphi}_{n,i}(r)
  \label{trddef2n}
  \\
  &=
  \sum_{i=1}^{N_{n}}
  \rho_{n,i}^{(g)}(r)
  \label{trddef3n}
  \\
  \rho_{p}^{(g)}(r)
  &\equiv
  \vec{\psi}_g^{\mathsf{T}}(r)
  \xi_{p}(r)
  \label{trddef1p}
  \\
  &=
  \sum_{i=N_n+1}^{N_{\text{tot}}}
  \left(
  \vec{\psi}_{g,X}^{\mathsf{T}}(r)
  +
  \vec{\psi}_{g,Y}^{\mathsf{T}}(r)
  \right)_i
  \tilde{\varphi}_{p,i}(r)
  \label{trddef2p}
  \\
  &=
  \sum_{i=N_n+1}^{N_{\text{tot}}}
  \rho_{p,i}^{(g)}(r).
  \label{trddef3p}
\end{align}
Consequently, the pole residue of the coordinate-space RPA response function $R_{qq'}(r, r'; E)$ factorizes into the direct product of the transition densities:
\begin{align}
  \operatorname{Res}\left[R_{qq'}(r,r'),E_g\right]  
  &=
  \pi
  \rho_{q}^{(g)}(r)
  \rho_{q'}^{(g)}(r').
  \label{eq:res_Rqq}
\end{align}

\subsection{The electromagnetic transition probability within Jost-RPA}

Above the particle emission threshold, continuous particle escape renders the effective Hamiltonian non-Hermitian, positioning the resonance poles $E_g$ in the fourth quadrant of the complex energy plane. In this continuum domain, matrix elements of physical observables are intrinsically complex quantities. Employing the multipole response function $R_F(E)$ formulated in Eq.~(\ref{eq:defRF}) of Appendix~\ref{app:jost_green}, the generalized complex electromagnetic transition probability $\widetilde{B}(EL)$ at the complex pole $E_g$ is defined via the contour residue:
\begin{eqnarray}
  \widetilde{B}(EL)
  \equiv
  \frac{1}{2i}
  \oint_{C_g} dE
  R_F(E)
  =
  \pi
  \operatorname{Res}
  \left[
    R_F, E_g
    \right].
  \label{eq:B_contour}
\end{eqnarray}
In the narrow-resonance limit ($\operatorname{Im} E_g \to 0^-$), the contour encircling the pole collapses onto the real energy axis, and $\widetilde{B}(EL)$ smoothly converges to the standard real-valued transition probability $B(EL)$.

By substituting the definition of the response function $R_F(E)$ [Eq.~(\ref{eq:defRF}) of Appendix~\ref{app:jost_green}] and the residue factorization of $R_{qq'}(r,r')$ [Eq.~(\ref{eq:res_Rqq})], the double radial integrals decouple, yielding the residue of $R_F(E)$ as the square of the total transition amplitude $\mathcal{F}_g^{(\tau)}$:
\begin{align}
  \operatorname{Res}\left[R_{F},E_g\right]  
  &=
  \mathcal{F}_g^{(\tau)2}
  \label{resRF}
\end{align}
where the transition amplitude $\mathcal{F}_g^{(\tau)}$ is defined by the spatial integral of the transition density $\rho_q^{(g)}(r)$ contracted with the external multipole operator $f_q^{(\tau)}(r)$:
\begin{align}
  \mathcal{F}_g^{(\tau)}
  &\equiv
  \int dr
  \sum_{q=n,p}
  f_{q}^{(\tau)}(r)
  \rho_{q}^{(g)}(r)
  \label{defFamp1}
  \\
  &=
  \int dr
  f_{n}^{(\tau)}(r)
  \sum_{i=1}^{N_{n}}
  \rho_{n,i}^{(g)}(r)
  \nonumber\\
  &+
  \int dr
  f_{p}^{(\tau)}(r)
  \sum_{i=N_n+1}^{N_{\text{tot}}}
  \rho_{p,i}^{(g)}(r)
  \\
  &\equiv
  \sum_{i=1}^{N_{\text{tot}}}
  F_{g,i}^{(\tau)}.
  \label{defFamp2}
\end{align}
with 
\begin{equation}
  F_{g,i}^{(\tau)}
  =
  \left\{
  \begin{array}{ll}
    \displaystyle{
      \int dr
      f_{n}^{(\tau)}(r)
      \rho_{n,i}^{(g)}(r)
      }
    &
    \text{(for $1\leq i\leq N_{n}$)}
    \\
    \equiv
    F_{g,i}^{(\tau,n)}
    &
    \\
    \\
    \displaystyle{
      \int dr
      f_{p}^{(\tau)}(r)
      \rho_{p,i}^{(g)}(r)
      }
    &
    \text{(for $N_{n}+1\leq i\leq N_{\text{tot}}$)}
    \\
    \equiv
    F_{g,i}^{(\tau,p)}
    &
  \end{array}
  \right.
\end{equation}
where $\rho_{q}^{(g)}(r)$ and $\rho_{q,i}^{(g)}(r)$ are the transition density and
its components at a pole $E_g$ defined by Eqs.~(\ref{trddef1n})--(\ref{trddef3p}).

\subsection{Coherence, Participation Rate and Transition Probability}
\label{subsec:coherence_ratio}
A collective state in the RPA framework is defined as a coherent superposition of
a large number of particle-hole excitations generated by the residual interaction. 
In order to mathematically formulate this concept and directly connect
its internal structure to $\widetilde{B}(EL)$, we define the phase coherence 
index $C_g^{(\tau)}$ and the participation rate $\eta_g^{(\tau)}$ as follows. 
More generally, the coherent superposition of multi-channel configurations
has been universally recognized as the origin of collectivity~\cite{Co2013}.
Within the Jost-RPA framework, the coherence index $C_g^{(\tau)}$ is defined as
\begin{equation}
  C_g^{(\tau)2}
  \equiv
  \frac{|\mathcal{F}_g^{(\tau)}|^2}
       {\displaystyle{\left(\sum_{i=1}^{N_{\text{tot}}}|F_{g,i}^{(\tau)}|\right)^2}}
  \label{Cindex}
\end{equation}
Since $\mathcal{F}_g^{(\tau)}$ and $F_{g,i}^{(\tau)}$ defined in Eqs.(\ref{defFamp1}) and
(\ref{defFamp2}) are complex numbers, they can be represented as
\begin{align}
  \mathcal{F}_g^{(\tau)}
  &=
  |\mathcal{F}_g^{(\tau)}|e^{i\Theta_{g}^{(\tau)}}
  \\
  F_{g,i}^{(\tau)}
  &=
  |F_{g,i}^{(\tau)}|e^{i\theta_i}
\end{align}
by introducing their arguments $\Theta_{g}^{(\tau)}$ and $\theta_i$.
The physical meaning of $F_{g,i}^{(\tau)}$ is the transition amplitude of the $i$-th particle-hole channel (where the index $i$ specifies both the hole state and the angular momentum of the continuum particle) to the resonance pole $E_g$.
If the phases of the transitions from all hole states are in phase—that is,
if the transitions are coherent—then
\begin{equation}
  \theta_1=\theta_2=...=\theta_{N_{\text{tot}}}\equiv\theta_0.
\end{equation}
Consequently, when the resonant pole state is perfectly coherent, it satisfies
\begin{equation}
  |\mathcal{F}_g^{(\tau)}|
  =
  \sum_{i=1}^{N_{\text{tot}}}
  |F_{g,i}^{(\tau)}|,   
\end{equation}
and this result indicates that $C_g^{(\tau)}=1$ for the coherent state.

To quantify the effective number of participating configurations, two main streams of definitions exist in the literature. The first is the conventional participation ratio, widely used in studies of Anderson localization, quantum chaos, and state-vector complexity, defined by the ratio of the squared $\ell_2$-norm to the squared $\ell_4$-norm of a state-vector amplitude distribution (e.g., $P = (\sum |c_i|^2)^2 / \sum |c_i|^4$)~\cite{Wegner1980,Armstrong2012,Borgonovi2016}. While this conventional definition effectively characterizes the spatial or configurational localization/delocalization of stationary wave functions, our primary objective here is to quantify the effective number of channel transition amplitudes contributing to an external one-body multipole response and to decouple this participation scale from complex phase interference. The second approach, established in information theory and signal processing, defines the effective number of contributing components using the ratio of the squared $\ell_1$-norm to the squared $\ell_2$-norm, expressed as $(\sum |a_i|)^2 / \sum |a_i|^2$~\cite{Hoyer2004,Hurley2009}.

In the present framework, giant resonances and continuum response functions are evaluated from the coherent superposition of complex transition amplitudes $F_{g,i}^{(\tau)}$. Adopting the $\ell_1/\ell_2$-norm formulation is naturally suited for this purpose, as it directly permits the transition strength residue $|\widetilde{B}(EL)|$ to be algebraically factorized into the exact product of the phase coherence index $C_g^{(\tau)2}$, the participation number $N_{\text{tot}}\eta_g^{(\tau)}$, and the average single-particle radial scale.

Regarding terminology, the quantity derived from the $\ell_4$-norm has often been called the "participation ratio" in the literature, although its numerical value represents an effective \emph{number} of participating states (ranging from $1$ to the total dimension $N_{\text{tot}}$). To avoid ambiguity, we distinguish between the "number" and the "rate" (or ratio). In our formulation, the unscaled quantity representing the effective number of components is referred to as the participation \emph{number}, which is bounded between $1$ (single-channel dominance) and $N_{\text{tot}}$ (equal participation). By dividing this participation number by the total number of configurations $N_{\text{tot}}$ to obtain a proper fraction bounded between $1/N_{\text{tot}}$ and $1$, we define the normalized participation rate $\eta_g^{(\tau)}$ as

\begin{equation}
  \eta_g^{(\tau)}
  \equiv
  \frac{\displaystyle{\left(\sum_{i=1}^{N_{\text{tot}}}|F_{g,i}^{(\tau)}|\right)^2}}
    {N_{\text{tot}}\displaystyle{\sum_{i=1}^{N_{\text{tot}}}|F_{g,i}^{(\tau)}|^2}}.
  \label{etaindex}
\end{equation}
$\widetilde{B}(EL)$ can be written as
\begin{equation}
  \widetilde{B}(EL)
  =
  \pi
  \left(
  C_g^{(\tau)2}
  N_{\text{tot}}
  \eta_g^{(\tau)}
  \right)
  \left(
  \sum_{i=1}^{N_{\text{tot}}}
  |F_{g,i}^{(\tau)}|^2
  \right)
  e^{2i\Theta_g^{(\tau)}}
  \label{Bexp}.
\end{equation}
Equation~(\ref{Bexp}) factorizes the transition strength $\widetilde{B}(EL)$ into the product of the coherence index $C_g^{(\tau)2}$, the total number of channels $N_{\text{tot}}$, the participation rate $\eta_g^{(\tau)}$, the sum of squared channel amplitudes $\sum_i |F_{g,i}^{(\tau)}|^2$, and the phase factor $e^{2i\Theta_g^{(\tau)}}$.

An algebraic feature of this factorization is the cancellation of the amplitude scale. By substituting the definitions of $C_g^{(\tau)2}$ and $\eta_g^{(\tau)}$ [Eqs.~(\ref{Cindex}) and (\ref{etaindex})] into Eq.~(\ref{Bexp}), the overall scale factor $\sum_i |F_{g,i}^{(\tau)}|^2$ is cancelled by the denominator of $\eta_g^{(\tau)}$, while $(\sum_i |F_{g,i}^{(\tau)}|)^2$ cancels between the denominator of $C_g^{(\tau)2}$ and the numerator of $\eta_g^{(\tau)}$. As a result, the product $C_g^{(\tau)2} N_{\text{tot}} \eta_g^{(\tau)}$ reduces to
\begin{equation}
  C_g^{(\tau)2} N_{\text{tot}} \eta_g^{(\tau)}
  =
  \frac{|\mathcal{F}_g^{(\tau)}|^2}{\displaystyle{\sum_{i=1}^{N_{\text{tot}}} |F_{g,i}^{(\tau)}|^2}}
  =
  \frac{\displaystyle{\left|\sum_{i=1}^{N_{\text{tot}}} F_{g,i}^{(\tau)}\right|^2}}{\displaystyle{\sum_{i=1}^{N_{\text{tot}}} |F_{g,i}^{(\tau)}|^2}}.
  \label{defCNeta_cancelled}
\end{equation}

From Eq.~(\ref{defCNeta_cancelled}), the product $C_g^{(\tau)2} N_{\text{tot}} \eta_g^{(\tau)}$ is scale-invariant under a uniform scaling of the channel amplitudes, $F_{g,i}^{(\tau)} \to \alpha F_{g,i}^{(\tau)}$ ($\alpha > 0$). Under this transformation, both $C_g^{(\tau)2}$ and $\eta_g^{(\tau)}$ remain unchanged, whereas $\sum_i |F_{g,i}^{(\tau)}|^2$ scales as $\alpha^2$.

This property shows that Eq.~(\ref{Bexp}) separates:
\begin{enumerate}[(i)]
  \item the coherent enhancement given by $C_g^{(\tau)2} N_{\text{tot}} \eta_g^{(\tau)}$, which is determined by the relative phases and the amplitude distribution among channels, from
  \item the overall scale of single-particle transition strengths, $\sum_i |F_{g,i}^{(\tau)}|^2$, which reflects the radial matrix elements of the transition operator.
\end{enumerate}

Consequently, a large value of $|\widetilde{B}(EL)|$ alone does not necessarily indicate a collective state: it can be driven solely by the single-particle scale $\sum_i |F_{g,i}^{(\tau)}|^2$, as for a single dominant channel (Sec.~\ref{subsec:bounds}), rather than by a coherent multi-channel enhancement. Dividing $|\widetilde{B}(EL)|$ by the amplitude scale $\sum_i |F_{g,i}^{(\tau)}|^2$ to evaluate $C_g^{(\tau)2} N_{\text{tot}} \eta_g^{(\tau)}$ is therefore necessary to properly assess collectivity.

\subsection{Bounds and Classification of Collective Modes}
\label{subsec:bounds}

To establish a quantitative criterion for collectivity, we consider the algebraic upper bound of the product $C_g^{(\tau)2} N_{\text{tot}} \eta_g^{(\tau)}$. Using the triangle inequality for the complex amplitudes $F_{g,i}^{(\tau)}$,
\begin{equation}
  \left|\sum_{i=1}^{N_{\text{tot}}} F_{g,i}^{(\tau)}\right|^2
  \le
  \left(\sum_{i=1}^{N_{\text{tot}}} |F_{g,i}^{(\tau)}|\right)^2,
\end{equation}
and applying the Cauchy-Schwarz inequality to the $\ell_1$- and $\ell_2$-norms,
\begin{equation}
  \left(\sum_{i=1}^{N_{\text{tot}}} |F_{g,i}^{(\tau)}|\right)^2
  \le
  N_{\text{tot}} \sum_{i=1}^{N_{\text{tot}}} |F_{g,i}^{(\tau)}|^2,
\end{equation}
we obtain the upper bound
\begin{equation}
  C_g^{(\tau)2} N_{\text{tot}} \eta_g^{(\tau)}
  =
  \frac{\displaystyle{\left|\sum_{i=1}^{N_{\text{tot}}} F_{g,i}^{(\tau)}\right|^2}}{\displaystyle{\sum_{i=1}^{N_{\text{tot}}} |F_{g,i}^{(\tau)}|^2}}
  \le
  N_{\text{tot}}.
  \label{CSbound}
\end{equation}
The theoretical maximum $N_{\text{tot}}$ corresponds to the ideal limit where all $N_{\text{tot}}$ channels contribute with identical amplitude and phase ($C_g^{(\tau)2} = 1$ and $\eta_g^{(\tau)} = 1$).

Physically, $C_g^{(\tau)2} N_{\text{tot}} \eta_g^{(\tau)}$ represents the \emph{effective number of coherent channels}.

In realistic nuclei, unperturbed single-particle energies and radial overlaps vary among channels, preventing the ideal limit $N_{\text{tot}}$ from being strictly reached. Nevertheless, the value of $C_g^{(\tau)2} N_{\text{tot}} \eta_g^{(\tau)}$ provides a systematic classification of excitation modes:
\begin{enumerate}[(i)]
  \item \textbf{Giant collective resonance}:
  Residual interactions coherently align the phases of numerous particle-hole channels ($C_g^{(\tau)2} \lesssim 1$) with broad participation ($\eta_g^{(\tau)} \lesssim 1$), resulting in
  \begin{equation}
    1 \ll C_g^{(\tau)2} N_{\text{tot}} \eta_g^{(\tau)} \lesssim N_{\text{tot}}.
  \end{equation}
  The coherent enhancement significantly exceeds the single-particle baseline of $1$, reaching the order of the total channel number $N_{\text{tot}}$.
  
  \item \textbf{Partial / subspace collective mode}:
  When the coherent excitation is physically confined to a subset of $N_{\text{sub}}$ channels ($N_{\text{sub}} \ll N_{\text{tot}}$), applying the Cauchy-Schwarz inequality to the active subspace yields the constrained upper bound
  \begin{equation}
    C_g^{(\tau)2} N_{\text{tot}} \eta_g^{(\tau)} \le N_{\text{sub}}.
    \label{Subspacebound}
  \end{equation}
  In such a mode, $1 < C_g^{(\tau)2} N_{\text{tot}} \eta_g^{(\tau)} \lesssim N_{\text{sub}}$, reflecting a collective enhancement localized within a restricted subspace (such as valence neutron excitations in the Pygmy Dipole Resonance). A dedicated microscopic analysis of these subspace modes, in conjunction with spatial transition densities, will be presented in a forthcoming publication.
  
  \item \textbf{Non-collective single-particle mode}:
  $\eta_g^{(\tau)} \approx 1/N_{\text{tot}}$ and $C_g^{(\tau)2} \approx 1$, leading to $C_g^{(\tau)2} N_{\text{tot}} \eta_g^{(\tau)} \approx 1$, independently of the nucleus and the excitation multipolarity. A single channel dominates the transition without multi-channel enhancement.
  
  \item \textbf{Phase cancellation among channels}:
  When many channels participate with significant amplitudes ($\eta_g^{(\tau)} \gg 1/N_{\text{tot}}$) but their phase differences lead to a cancellation in the cross terms ($C_g^{(\tau)2} \ll 1/N_{\text{tot}}$), the product satisfies $C_g^{(\tau)2} N_{\text{tot}} \eta_g^{(\tau)} \ll 1$. In this case, the total transition strength $|\widetilde{B}(EL)|$ is suppressed despite the multi-channel participation.
\end{enumerate}

We emphasize that the collectivity metric $C_g^{(\tau)2} N_{\text{tot}} \eta_g^{(\tau)}$ is formulated with respect to the physically specified particle-hole continuum channel space and serves as an internal microscopic diagnostic tailored to continuum RPA calculations, rather than a basis-invariant many-body operator expectation value. In particular, the baseline value $C_g^{(\tau)2} N_{\text{tot}} \eta_g^{(\tau)} = 1$ signifies that the continuum response is effectively dominated by a single particle-hole channel (a quasi-single-particle excitation mode) rather than implying a strictly non-interacting single-nucleon state.

\subsection{Branching ratios and partial decay widths}
\label{subsec:branching_ratio}
While the coherence index and participation rate in Sec.~\ref{subsec:coherence_ratio} quantify the internal collective structure under external multipole fields, the Takagi factorization of the open-channel residue $\operatorname{Res}\left[\mathcal{S}_{cc},E_g\right]$ [Eq.~(\ref{takagi2})] uniquely determines the continuum particle decay dynamics of the resonant state.

The Takagi factorization of Eq.~(\ref{takagi2}), $\operatorname{Res}[\mathcal{S}_{cc},E_g]=\sigma_g^{(c)}\vec{\omega}_g^{(c)}\vec{\omega}_g^{(c)\mathsf{T}}$ with $\|\vec{\omega}_g^{(c)}\|=1$, separates the open-channel residue into an overall scale, the singular value $\sigma_g^{(c)}$, and a normalized channel vector $\vec{\omega}_g^{(c)}$. As noted in Sec.~\ref{subsec:takagi_residue}, $\vec{\omega}_g^{(c)}$ is equivalent to the asymptotic partial-width amplitudes of the Gamow state of the pole \cite{Mizuyama2026_Takagi}, so that its components describe how the decay of the resonance is shared among the open channels $i \in \mathcal{C}$. We therefore call $\sigma_g^{(c)}$ the residue width. Because the residue $A \equiv \operatorname{Res}[\mathcal{S}_{cc},E_g]$ is a complex symmetric matrix of rank one, $A^\dagger A = \sigma_g^{(c)2}\vec{\omega}_g^{(c)*}\vec{\omega}_g^{(c)\mathsf{T}}$ has the single nonzero eigenvalue $\sigma_g^{(c)2}$, and the residue width equals the norm of the residue matrix, whether the Frobenius norm or the spectral norm is used:
\begin{equation}
  \begin{split}
  \sigma_g^{(c)} &= \|A\|_{\text{F}} \equiv \Bigl(\sum_{i,j=1}^{c}|A_{ij}|^2\Bigr)^{1/2} \\
  &= \|A\|_2 \equiv \max_{\|\vec{x}\|=1}\|A\vec{x}\| .
  \end{split}
  \label{eq:sg_c_gamma}
\end{equation}
We define the partial decay widths $\Gamma_i$ and the relative branching ratios $\text{br}_i$ (or branching weights) by the components of the Takagi vector:
\begin{align}
  \Gamma_i &= \sigma_g^{(c)} |\omega_{g,i}^{(c)}|^2,
  \label{eq:gamma_i_takagi}
  \\
  \text{br}_i &\equiv |\omega_{g,i}^{(c)}|^2 = \frac{\Gamma_i}{\sum_{j=1}^c \Gamma_j},
  \label{eq:br_i_takagi}
\end{align}
which satisfy the exact sum rule $\sum_{i=1}^{c} \text{br}_i = 1$.
These definitions use only the computed residue and require no parametrization of $\mathcal{S}_{cc}(E)$ near the pole. Because $\vec{\omega}_g^{(c)}$ is normalized, $\text{br}_i$ does not depend on the overall scale $\sigma_g^{(c)}$ of the residue; how this scale is related to the pole width $\Gamma_g$ is derived below.

\begin{table*}[htbp]
\centering
\caption{Distinct physical definitions of resonance widths in open continuum systems and their mathematical realizations in the present Jost-RPA framework.}
\label{tab:width_definitions}
\begin{ruledtabular}
\begin{tabular}{lll}
Width Quantity & Mathematical Definition & Physical Meaning / Role \\
\hline
\textbf{Pole Width} $\Gamma_g$ & $E_g = \operatorname{Re}E_g - i\Gamma_g/2$ & Intrinsic resonance lifetime ($\tau = \hbar/\Gamma_g$) \\
\textbf{Residue Width} $\sigma_g^{(c)}$ & $\|\operatorname{Res}[\mathcal{S}_{cc}, E_g]\|$ (Takagi singular value) & Residue scale; $=\Gamma_g$ if the rest of $\mathcal{S}_{cc}$ is constant across the pole \\
\hline
\textbf{Branching Ratio} $\mathrm{br}_i$ & $|\omega_{g,i}^{(c)}|^2 = \Gamma_i/\sum_j \Gamma_j$ & Relative decay distribution over the open channels (independent of $\sigma_g^{(c)}$) \\
\end{tabular}
\end{ruledtabular}
\end{table*}

Table~\ref{tab:width_definitions} summarizes the width quantities used in this work: (i) the \emph{pole width} $\Gamma_g$ of $E_g=\operatorname{Re}E_g-i\Gamma_g/2$, a constant in the complex energy plane that defines the intrinsic quantum-mechanical lifetime $\tau=\hbar/\Gamma_g$; (ii) the \emph{residue width} $\sigma_g^{(c)}$ of Eq.~(\ref{eq:sg_c_gamma}), which sets the overall scale of the residue and thus of the on-shell transition strength near the pole; and (iii) the relative branching ratios $\text{br}_i$ of Eq.~(\ref{eq:br_i_takagi}).

We now derive the relation between $\sigma_g^{(c)}$ and $\Gamma_g$. Closed channels decay exponentially at large distances and carry no asymptotic flux. The open channels therefore carry the entire flux, and on the real-energy interval of the Riemann sheet the open-channel S-matrix is unitary, $\mathcal{S}_{cc}(E)\mathcal{S}_{cc}^\dagger(E)=\mathbf{1}$, irrespective of the coupling to closed configurations. The decay of the resonance at the rate $\Gamma_g$ thus proceeds entirely through the open channels, and for an isolated narrow pole on a background that varies slowly across its width one expects $\sigma_g^{(c)} \approx \Gamma_g$. In general, however, $\sigma_g^{(c)}/\Gamma_g$ deviates from unity, and the deviation follows exactly from the symmetry and the unitarity of $\mathcal{S}_{cc}$.

Because $\mathcal{S}_{cc}$ is symmetric, $\mathcal{S}_{cc}^\dagger=\mathcal{S}_{cc}^*$, and the unitarity reads $\mathcal{S}_{cc}^{-1}(E)=[\mathcal{S}_{cc}(E)]^*$ on the real axis. Its analytic continuation, $\mathcal{S}_{cc}^{-1}(E)=[\mathcal{S}_{cc}(E^*)]^*$, is analytic at $E=E_g$, because the upper half of the Riemann sheet coincides with the physical sheet, where $\mathcal{S}_{cc}$ has no pole at $E_g^*$. We separate the pole term of $g$ from $\mathcal{S}_{cc}$,
\begin{equation}
  \mathcal{S}_{cc}(E)=\mathcal{S}_{\text{B},g}(E)+\frac{\sigma_g^{(c)}\vec{\omega}_g^{(c)}\vec{\omega}_g^{(c)\mathsf{T}}}{E-E_g},
  \label{eq:Scc_decomp}
\end{equation}
where $\mathcal{S}_{\text{B},g}$ contains everything except the pole term of $g$, including the other poles, and is analytic at $E_g$ and $E_g^*$. The product $\mathcal{S}_{cc}^{-1}(E)\mathcal{S}_{cc}(E)=\mathbf{1}$ has no pole at $E_g$. By Eq.~(\ref{eq:Scc_decomp}), its residue there is $\sigma_g^{(c)}\mathcal{S}_{cc}^{-1}(E_g)\vec{\omega}_g^{(c)}\vec{\omega}_g^{(c)\mathsf{T}}$, which must therefore vanish. Hence $\mathcal{S}_{cc}^{-1}(E_g)\vec{\omega}_g^{(c)}=0$, and the complex conjugate of this relation, with the analytic continuation above, gives $\mathcal{S}_{cc}(E_g^*)\vec{\omega}_g^{(c)*}=0$. Inserting Eq.~(\ref{eq:Scc_decomp}) at $E=E_g^*$, where $E_g^*-E_g=i\Gamma_g$, and using $\vec{\omega}_g^{(c)\mathsf{T}}\vec{\omega}_g^{(c)*}=\|\vec{\omega}_g^{(c)}\|^2=1$, we obtain
\begin{equation}
  \mathcal{S}_{\text{B},g}(E_g^*)\,\vec{\omega}_g^{(c)*}=i\,\frac{\sigma_g^{(c)}}{\Gamma_g}\,\vec{\omega}_g^{(c)}.
  \label{eq:SB_omega}
\end{equation}
Taking the norm of both sides gives the exact relation
\begin{equation}
  \frac{\sigma_g^{(c)}}{\Gamma_g}=\bigl\|\mathcal{S}_{\text{B},g}(E_g^*)\,\vec{\omega}_g^{(c)*}\bigr\| .
  \label{eq:R_exact}
\end{equation}

Equation~(\ref{eq:R_exact}) evaluates the non-pole part at the complex energy $E_g^*$. To relate it to the energy dependence of $\mathcal{S}_{\text{B},g}$ on the real axis, we expand $\mathcal{S}_{\text{B},g}$ to first order over the distance $\Gamma_g/2$ between $\operatorname{Re}E_g$ and $E_g^*=\operatorname{Re}E_g+i\Gamma_g/2$:
\begin{equation}
  \mathcal{S}_{\text{B},g}(E_g^*) \simeq \mathcal{S}_{\text{B},g}(\operatorname{Re}E_g) + \frac{i\Gamma_g}{2}\frac{d\mathcal{S}_{\text{B},g}}{dE}(\operatorname{Re}E_g).
  \label{eq:SB_expand}
\end{equation}
At the real energy $\operatorname{Re}E_g$, where $\operatorname{Re}E_g-E_g=i\Gamma_g/2$, Eq.~(\ref{eq:Scc_decomp}) gives
\begin{equation}
  \mathcal{S}_{\text{B},g}(\operatorname{Re}E_g) = \mathcal{S}_{cc}(\operatorname{Re}E_g) + \frac{2i\sigma_g^{(c)}}{\Gamma_g}\,\vec{\omega}_g^{(c)}\vec{\omega}_g^{(c)\mathsf{T}}.
  \label{eq:SB_real}
\end{equation}
Substituting Eqs.~(\ref{eq:SB_expand}) and (\ref{eq:SB_real}) into Eq.~(\ref{eq:SB_omega}) and using $\vec{\omega}_g^{(c)\mathsf{T}}\vec{\omega}_g^{(c)*}=1$ again yields
\begin{equation}
  \begin{split}
  \mathcal{S}_{cc}(\operatorname{Re}E_g)\,\vec{\omega}_g^{(c)*} \simeq -i\biggl[ &\frac{\sigma_g^{(c)}}{\Gamma_g}\,\vec{\omega}_g^{(c)} \\
  &+ \frac{\Gamma_g}{2}\frac{d\mathcal{S}_{\text{B},g}}{dE}(\operatorname{Re}E_g)\,\vec{\omega}_g^{(c)*} \biggr].
  \end{split}
  \label{eq:S0_omega}
\end{equation}
Since $\mathcal{S}_{cc}(\operatorname{Re}E_g)$ is unitary, the left-hand side has unit norm. For the right-hand side, with $a=\sigma_g^{(c)}/\Gamma_g$, $\epsilon=\Gamma_g/2$ and $\vec{v}=(d\mathcal{S}_{\text{B},g}/dE)(\operatorname{Re}E_g)\,\vec{\omega}_g^{(c)*}$,
\begin{equation}
  \begin{split}
  \|a\vec{\omega}_g^{(c)}+\epsilon\vec{v}\| &= \bigl[a^2+2a\epsilon\operatorname{Re}(\vec{\omega}_g^{(c)\dagger}\vec{v})+\epsilon^2\|\vec{v}\|^2\bigr]^{1/2} \\
  &\simeq a+\epsilon\operatorname{Re}(\vec{\omega}_g^{(c)\dagger}\vec{v})
  \end{split}
  \label{eq:norm_expand}
\end{equation}
to first order in $\epsilon$. Setting this equal to unity gives
\begin{equation}
  \frac{\sigma_g^{(c)}}{\Gamma_g}\simeq1-\frac{\Gamma_g}{2}\operatorname{Re}\Bigl[\vec{\omega}_g^{(c)\dagger}\,\frac{d\mathcal{S}_{\text{B},g}}{dE}(\operatorname{Re}E_g)\,\vec{\omega}_g^{(c)*}\Bigr].
  \label{eq:R_first}
\end{equation}
Hence $\sigma_g^{(c)}=\Gamma_g$ when the rest of $\mathcal{S}_{cc}$ does not vary across the pole, and deviations of either sign arise, in proportion to the pole width, from its energy variation along the decay direction. In particular, $\sigma_g^{(c)}<\Gamma_g$ does not mean that part of the width is lost to closed channels: the whole decay flux is carried by the open channels, and closed channels, thresholds and barriers affect $\sigma_g^{(c)}/\Gamma_g$ only through the energy dependence of $\mathcal{S}_{\text{B},g}$ that they cause. Accordingly, $\sigma_g^{(c)}$ and $\Gamma_i=\sigma_g^{(c)}|\omega_{g,i}^{(c)}|^2$ are residue widths, which coincide with the decay rates in the narrow isolated limit.

One source of such an energy variation is provided by centrifugal and Coulomb barriers. Near the thresholds, the barrier penetrabilities make the open-channel amplitudes, and hence $\mathcal{S}_{\text{B},g}$, vary steeply with energy; in the Feshbach projection-operator theory \cite{Feshbach1958, Feshbach1962} and $R$-matrix theory \cite{Lane1958}, the same effect appears as the energy dependence of the self-energy. The neighboring poles, which are part of $\mathcal{S}_{\text{B},g}$, contribute in addition. Equation~(\ref{eq:R_first}) contains all these contributions without separating them. They determine how $\sigma_g^{(c)}$ differs from $\Gamma_g$.

\section{Numerical results}
\label{sec:results}

In this section, we apply the present theoretical framework to investigate the collectivity indices, channel participation rates, and decay branching ratios of nuclear continuum resonances. 
The single-particle Woods-Saxon mean-field potentials [Eqs.~(\ref{eq:vcent})--(\ref{eq:vcoul}) of Appendix~\ref{app:jost_formalism}] and the Skyrme-type residual particle-hole interaction [Eqs.~(\ref{eq:vqq})--(\ref{eq:vqqbar}) of Appendix~\ref{app:jost_formalism}] are evaluated using the identical parameter sets adopted in the previous JRPA studies \cite{Mizuyama2023_JRPA1, Mizuyama2024_Eigenphase, Mizuyama2025_JRPA2}. Specifically, we adopt the Woods-Saxon parameters $V_0 = 60\text{ MeV}$, $V_{\text{ls}} = 15\text{ MeV}$, $r_0 = r_{\text{ls}} = r_{\text{c}} = 1.2\text{ fm}$, $a_0 = a_{\text{ls}} = 0.65\text{ fm}$, and $\kappa = 0.67$, along with the Skyrme interaction parameters $t_0 = -1100\text{ MeV fm}^3$, $t_3 = 16000\text{ MeV fm}^6$, $x_0 = 0.5$, and $x_3 = 1.0$.

While earlier JRPA calculations \cite{Mizuyama2023_JRPA1, Mizuyama2024_Eigenphase, Mizuyama2025_JRPA2} employed the standard Numerov algorithm \cite{Numerov1924}, the present calculations utilize the multichannel Renormalized Numerov method \cite{Johnson1977} coupled with the Contour Integral Root-finding (CIRR) method \cite{Kravanja2000, Sakurai2003} for efficient and stable complex domain searches.

\subsection{Isovector dipole ($L=1, \tau=1$) continuum response}
\label{subsec:O16_L1_results}

\begin{table*}[htbp]
\centering
\caption{Complex resonance pole energies $E_g = \operatorname{Re}(E_g) + i\operatorname{Im}(E_g)$, open-channel Takagi singular values $\sigma_g^{(c)}$, collectivity parameters ($C_g^{(1)2}, \Theta_g^{(1)}/\pi, N_{\text{tot}}\eta_g^{(1)}, C_g^{(1)2} N_{\text{tot}}\eta_g^{(1)}$ with $N_{\text{tot}} = 14$), and complex transition strengths $\operatorname{Re}[\widetilde{B}(E1)]/\pi$, $\operatorname{Im}[\widetilde{B}(E1)]/\pi$ evaluated via Eq.~(\ref{eq:B_contour}) for the isovector dipole ($L=1, \tau=1$) resonance poles in $^{16}\text{O}$ on Riemann sheets 4 and 5 ($\text{RS}=4, 5$).}
\label{tab:O16_E1_collectivity}
\begin{ruledtabular}
\begin{tabular}{ccrrrrrrrrr}
\multirow{2}{*}{RS} & Pole & $\operatorname{Re}(E_g)$ & $\operatorname{Im}(E_g)$ & $\sigma_g^{(c)}$ & \multicolumn{4}{c}{Collectivity Parameters ($\tau=1$)} & \multicolumn{2}{c}{$\widetilde{B}(E1)/\pi$ [$e^2\text{fm}^2$]} \\
\cline{6-9} \cline{10-11}
 & No. & [MeV] & [MeV] & [MeV] & $C_g^{(1)2}$ & $\Theta_g^{(1)}/\pi$ & $N_{\text{tot}}\eta_g^{(1)}$ & $C_g^{(1)2} N_{\text{tot}}\eta_g^{(1)}$ & $\operatorname{Re}(\widetilde{B})/\pi$ & $\operatorname{Im}(\widetilde{B})/\pi$ \\
\hline
\multirow{8}{*}{4} 
 & 1 & 17.126 & $-0.021$ & 0.041 & 0.0006 & $+0.073$ & 5.417 & 0.003 & $1.67 \times 10^{-4}$ & $8.22 \times 10^{-5}$ \\
 & 2 & 17.412 & $-1.840$ & 0.752 & 0.1631 & $-0.305$ & 7.687 & 1.253 & $-3.14 \times 10^{-3}$ & $-8.66 \times 10^{-3}$ \\
 & 3 & 18.528 & $-0.719$ & 0.368 & 0.0580 & $+0.809$ & 8.542 & 0.495 & $1.73 \times 10^{-3}$ & $-4.41 \times 10^{-3}$ \\
 & 4 & 19.294 & $-0.954$ & 0.481 & 0.1971 & $+0.762$ & 10.319 & 2.034 & $4.13 \times 10^{-4}$ & $-5.48 \times 10^{-3}$ \\
 & 5 & 19.349 & $-0.529$ & 1.112 & 0.0654 & $-0.969$ & 9.084 & 0.594 & $9.28 \times 10^{-2}$ & $1.83 \times 10^{-2}$ \\
 & 6 & 19.392 & $-0.977$ & 0.411 & 0.1106 & $+0.234$ & 10.928 & 1.208 & $1.20 \times 10^{-3}$ & $1.17 \times 10^{-2}$ \\
 & 7 & 20.765 & $-0.344$ & 0.815 & 0.1385 & $-0.958$ & 10.851 & 1.503 & $5.79 \times 10^{-2}$ & $1.56 \times 10^{-2}$ \\
 & 8 & 21.245 & $-1.713$ & 0.799 & 0.2841 & $-0.767$ & 11.804 & 3.353 & $3.34 \times 10^{-3}$ & $3.06 \times 10^{-2}$ \\
\hline
5 & 1 & 21.718 & $-0.624$ & 1.565 & 0.2931 & $-0.221$ & 11.915 & 3.492 & $2.22 \times 10^{-2}$ & $-1.22 \times 10^{-1}$ \\
\end{tabular}
\end{ruledtabular}
\end{table*}

\begin{figure}[htbp]
\centering
\includegraphics[width=\columnwidth]{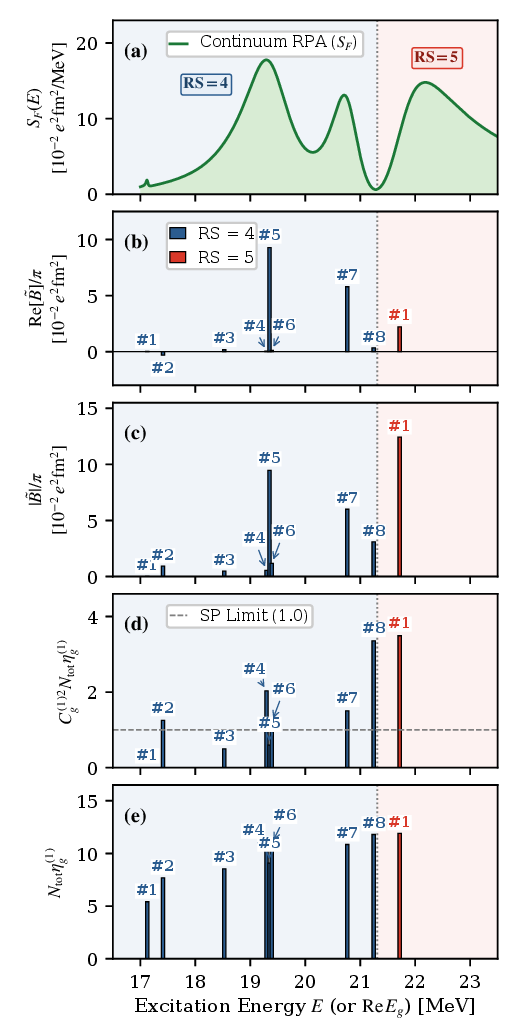}
\caption{(Color online) Microscopic decomposition of the isovector dipole ($E1, \tau=1$) continuum response in $^{16}\text{O}$ across Riemann sheets 4 and 5 ($\text{RS}=4, 5$). (a) Total continuum RPA strength function $S_F(E)$ (green solid curve with green shading). (b) Real part of the complex transition strength $\operatorname{Re}[\widetilde{B}(E1)]/\pi$ evaluated at the resonance poles $E_g$ (blue bars for $\text{RS}=4$ and red bar for $\text{RS}=5$). (c) Modulus of the transition strength $|\widetilde{B}(E1)|/\pi$. (d) Collectivity enhancement factor $C_g^{(1)2} N_{\text{tot}}\eta_g^{(1)}$ with the horizontal dashed line indicating the single-particle limit ($=1.0$). (e) Raw channel coherence sum $N_{\text{tot}}\eta_g^{(1)}$. Shaded background regions distinguish $\text{RS}=4$ ($E \le 21.31\text{ MeV}$, light blue) and $\text{RS}=5$ ($E \ge 21.31\text{ MeV}$, light red), separated by the particle decay threshold at $E_{\text{th}} = 21.31\text{ MeV}$ (vertical dotted line).}
\label{fig:O16_E1_analysis}
\end{figure}

Figure~\ref{fig:O16_E1_analysis} and Table~\ref{tab:O16_E1_collectivity} present the microscopic decomposition of the isovector dipole ($E1, \tau=1$) continuum response in $^{16}\text{O}$. 
Owing to the CIRR domain search, Table~\ref{tab:O16_E1_collectivity} includes several resonance poles that were not identified in earlier JRPA studies \cite{Mizuyama2023_JRPA1, Mizuyama2024_Eigenphase, Mizuyama2025_JRPA2}. In particular, in the vicinity of $19.3\text{ MeV}$, the present calculation resolves a closely spaced triplet of poles: Pole~\#4 ($19.29 - 0.95i\text{ MeV}$), Pole~\#5 ($19.35 - 0.53i\text{ MeV}$), and Pole~\#6 ($19.39 - 0.98i\text{ MeV}$), separated by only tens of keV. In earlier calculations based on the standard Numerov method, Poles~\#4 and \#6 were obscured by the dominant Pole~\#5 due to their weak transition strength and numerical precision limits in the closed channels. Because the single-particle potential and interaction parameters are identical to those of Refs.~\cite{Mizuyama2023_JRPA1, Mizuyama2024_Eigenphase, Mizuyama2025_JRPA2}, these newly identified poles represent genuine continuum RPA eigensolutions.

A prominent feature revealed by this complete spectrum is the clear separation between single-particle dominated peaks and collective multi-channel resonances. 
On Riemann sheet 4 ($\text{RS}=4$, $E \le 21.31\text{ MeV}$), Pole~\#5 at $E_g = 19.35 - 0.53i\text{ MeV}$ provides the largest real transition strength $\operatorname{Re}[\widetilde{B}(E1)]/\pi = 9.28 \times 10^{-2}\ e^2\text{fm}^2$, corresponding to the dominant narrow peak in $S_F(E)$ shown in Fig.~\ref{fig:O16_E1_analysis}(a). However, its collectivity index $C_g^{(1)2} N_{\text{tot}}\eta_g^{(1)} = 0.594$ is below unity [Fig.~\ref{fig:O16_E1_analysis}(d)], demonstrating that this peak is predominantly driven by a single unperturbed particle-hole transition with a large single-particle matrix element rather than collective multi-channel superposition.

In contrast, Pole~\#1 on Riemann sheet 5 ($\text{RS}=5$, $E_g = 21.72 - 0.62i\text{ MeV}$) represents the primary collective isovector giant dipole resonance (IVGDR). As seen in Figs.~\ref{fig:O16_E1_analysis}(c)--(e), this state exhibits the maximum intrinsic transition strength scale $|\widetilde{B}(E1)|/\pi = 0.124\ e^2\text{fm}^2$, the maximum channel coherence $N_{\text{tot}}\eta_g^{(1)} = 11.915$, and a pronounced collectivity factor $C_g^{(1)2} N_{\text{tot}}\eta_g^{(1)} = 3.492 \gg 1$. Because this state possesses a non-zero rotation phase $\Theta_g^{(1)}/\pi = -0.221$, its large imaginary residue produces significant asymmetric resonance-continuum interference in the real-axis strength function $S_F(E)$. 

It is instructive to contrast the collective IVGDR state with the neighboring resonance poles on Riemann sheet 4, particularly Pole~\#4 ($19.29 - 0.95i\text{ MeV}$) and Pole~\#6 ($19.39 - 0.98i\text{ MeV}$). Although both states involve the participation of numerous channels ($N_{\text{tot}}\eta_g^{(1)} \approx 10.32$ and $10.93$, respectively), their complex rotation phases ($\Theta_g^{(1)}/\pi = +0.762$ and $+0.234$) produce destructive interference among different particle-hole transitions. This severe phase cancellation quenches the transition strengths down to $|\widetilde{B}(E1)|/\pi \approx 5.48 \times 10^{-3}\ e^2\text{fm}^2$ and $1.18 \times 10^{-2}\ e^2\text{fm}^2$, corresponding to the phase-cancellation regime classified in Sec.~\ref{subsec:coherence_ratio}.

\begin{table*}[htbp]
\centering
\caption{Partial decay widths $\Gamma_i$ [MeV] and relative branching ratios $\mathrm{br}_i$ [\%] evaluated via open-channel Takagi factorization [Eqs.~(\ref{eq:gamma_i_takagi})--(\ref{eq:br_i_takagi})] for the three prominent resonance poles in $^{16}\text{O}$ ($L=1, \tau=1$): the dominant single-particle resonance on $\text{RS}=4$ (Pole~\#5, $E_g = 19.35 - 0.53i\text{ MeV}$), the pre-threshold transition resonance on $\text{RS}=4$ (Pole~\#7, $E_g = 20.76 - 0.34i\text{ MeV}$), and the collective IVGDR giant resonance on $\text{RS}=5$ (Pole~\#1, $E_g = 21.72 - 0.62i\text{ MeV}$). Dashes (---) indicate energetically closed decay channels on the respective Riemann sheet.}
\label{tab:O16_E1_branching}
\begin{ruledtabular}
\begin{tabular}{ccrrrrrr}
\makecell[c]{Nucleon\\Type} & \makecell[c]{Decay Channel\\(Hole $\to$ Particle Wave)} & \multicolumn{2}{c}{RS=4, Pole \#5 ($19.35\text{ MeV}$)} & \multicolumn{2}{c}{RS=4, Pole \#7 ($20.76\text{MeV}$)} & \multicolumn{2}{c}{RS=5, Pole \#1 ($21.72\text{ MeV}$)} \\
\cline{3-4} \cline{5-6} \cline{7-8}
 & & $\Gamma_i$ [MeV] & $\mathrm{br}_i$ [\%] & $\Gamma_i$ [MeV] & $\mathrm{br}_i$ [\%] & $\Gamma_i$ [MeV] & $\mathrm{br}_i$ [\%] \\
\hline
\multirow{5}{*}{Neutron} 
 & $1p_{1/2} \to s_{1/2}\ (\Delta l = -1)$ & 0.0247 &  2.22 & 0.0594 &  7.28 & 0.0534 &  3.41 \\
 & $1p_{1/2} \to d_{3/2}\ (\Delta l = +1)$ & 0.1603 & 14.41 & 0.2205 & 27.06 & 0.2995 & 19.13 \\
 & $1p_{3/2} \to d_{5/2}\ (\Delta l = +1)$ & \multicolumn{2}{c}{---} & \multicolumn{2}{c}{---} & 0.0090 &  0.57 \\
 & $1p_{3/2} \to s_{1/2}\ (\Delta l = -1)$ & \multicolumn{2}{c}{---} & \multicolumn{2}{c}{---} & 0.0347 &  2.22 \\
 & $1p_{3/2} \to d_{3/2}\ (\Delta l = +1)$ & \multicolumn{2}{c}{---} & \multicolumn{2}{c}{---} & 0.0553 &  3.53 \\
\hline
\multirow{5}{*}{Proton} 
 & $1p_{1/2} \to s_{1/2}\ (\Delta l = -1)$ & 0.0838 &  7.53 & 0.0126 &  1.54 & 0.0831 &  5.31 \\
 & $1p_{1/2} \to d_{3/2}\ (\Delta l = +1)$ & 0.5038 & 45.29 & 0.1246 & 15.29 & 0.4469 & 28.55 \\
 & $1p_{3/2} \to d_{5/2}\ (\Delta l = +1)$ & 0.0792 &  7.12 & 0.0237 &  2.91 & 0.1611 & 10.29 \\
 & $1p_{3/2} \to s_{1/2}\ (\Delta l = -1)$ & 0.0966 &  8.69 & 0.0208 &  2.55 & 0.1798 & 11.48 \\
 & $1p_{3/2} \to d_{3/2}\ (\Delta l = +1)$ & 0.1639 & 14.74 & 0.3533 & 43.36 & 0.2425 & 15.50 \\
\hline
\multicolumn{2}{l}{Total Open Width $\sigma_g^{(c)}$ [MeV]} & \multicolumn{2}{c}{1.1124} & \multicolumn{2}{c}{0.8149} & \multicolumn{2}{c}{1.5653} \\
\multicolumn{2}{l}{Total Pole Width $\Gamma_g$ [MeV]} & \multicolumn{2}{c}{1.0584} & \multicolumn{2}{c}{0.6873} & \multicolumn{2}{c}{1.2487} \\
\multicolumn{2}{l}{Ratio $\sigma_g^{(c)} / \Gamma_g$} & \multicolumn{2}{c}{105.1\%} & \multicolumn{2}{c}{118.6\%} & \multicolumn{2}{c}{125.4\%} \\
\multicolumn{2}{l}{Neutron / Proton Sum [\%]} & \multicolumn{2}{c}{16.6\% / 83.4\%} & \multicolumn{2}{c}{34.3\% / 65.7\%} & \multicolumn{2}{c}{28.9\% / 71.1\%} \\
\end{tabular}
\end{ruledtabular}
\end{table*}

To elucidate the decay mechanisms of these continuum excitations, Table~\ref{tab:O16_E1_branching} displays the partial decay widths $\Gamma_i$ and relative branching ratios $\mathrm{br}_i$ for the three prominent resonance poles associated with the distinct structures in $S_F(E)$.
For the single-particle dominant resonance (RS=4, Pole~\#5 at $19.35\text{ MeV}$), the decay is concentrated into the proton channel from the valence hole state, with the $p(1p_{1/2} \to d_{3/2})$ mode alone accounting for $45.29\%$ of the total decay width, while the total proton emission reaches $83.4\%$.
As the excitation energy approaches the second peak (RS=4, Pole~\#7 at $20.76\text{ MeV}$), the decay pattern undergoes a substantial reorganization: the dominant decay path shifts to the deeper hole transition $p(1p_{3/2} \to d_{3/2})$ ($43.36\%$) along with a heightened neutron emission $n(1p_{1/2} \to d_{3/2})$ ($27.06\%$), reflecting the progressive activation of deeper-lying particle-hole configurations.

Finally, for the collective IVGDR state (RS=5, Pole~\#1 at $21.72\text{ MeV}$), all 10 open channels participate concurrently in the decay process. The individual branching ratios are broadly distributed across multiple neutron ($28.9\%$) and proton ($71.1\%$) channels, with no individual channel exceeding $28.6\%$. This multi-channel fragmentation of the decay width provides a clear microscopic indication of the coherent superimposition of all available particle-hole decay pathways in the continuum giant resonance.

 The microscopic decay analysis also characterizes Pole~\#4 and Pole~\#6, which do not manifest as observable peaks in the continuum strength function. For both poles the residue width is much smaller than the pole width ($\sigma_g^{(c)}/\Gamma_g = 25.18\%$ and $21.03\%$, respectively), which by Eq.~(\ref{eq:R_first}) reflects a strong energy variation of the rest of $\mathcal{S}_{cc}$ across these broad poles lying within tens of keV of Pole~\#5. Moreover, their open-channel decay is strongly localized into single proton emission pathways ($p(1p_{3/2} \to d_{5/2})$ carrying $64.93\%$ for Pole~\#4, and $p(1p_{3/2} \to d_{3/2})$ carrying $50.02\%$ for Pole~\#6) with virtually vanishing neutron decay ($<6\%$). This localized, non-collective decay behavior sharply contrasts with the balanced multi-channel fragmentation observed in the true IVGDR state (RS=5, Pole~\#1).

In addition, the evolution across the Riemann sheet boundary at $E_{\text{th}} = 21.31\text{ MeV}$ (the threshold for opening the neutron $1p_{3/2}^{-1}$ channels) reveals an important dynamical phenomenon regarding the pre-threshold onset of collectivity. Directly below the threshold, Pole~\#8 on $\text{RS}=4$ ($E_g = 21.25 - 1.71i\text{ MeV}$) does not form a discernible peak in the real-axis strength function $S_F(E)$ nor in the real transition strength $\operatorname{Re}[\widetilde{B}(E1)]/\pi = 3.34 \times 10^{-3}\ e^2\text{fm}^2$. To understand why this highly collective state is hidden from the real-axis spectrum, we examine the single-pole decomposition of the strength function:
\begin{align}
S_F^{(g)}(E) &= {-}\frac{1}{\pi} \operatorname{Im}\left[ \frac{\widetilde{B}_g}{E - E_g} \right] \nonumber \\
&= \frac{\operatorname{Re}[\widetilde{B}_g] (\Gamma_g/2) {-} \operatorname{Im}[\widetilde{B}_g] (E - \operatorname{Re}E_g)}{\pi \left[ (E - \operatorname{Re}E_g)^2 + (\Gamma_g/2)^2 \right]},
\label{eq:SF_pole_decomp}
\end{align}
where the complex transition strength $\widetilde{B}_g = |\widetilde{B}_g| e^{2i\Theta_g}$ (since $\widetilde{B}_g=\pi\mathcal{F}_g^{(\tau)2}$ and $\Theta_g=\arg\mathcal{F}_g^{(\tau)}$) is partitioned into a symmetric Lorentzian Breit-Wigner component $\operatorname{Re}[\widetilde{B}_g] = |\widetilde{B}_g| \cos2\Theta_g$ and an asymmetric dispersive interference component $\operatorname{Im}[\widetilde{B}_g] = |\widetilde{B}_g| \sin2\Theta_g$. 
Crucially, the large total width ($\Gamma_g = 3.43\text{ MeV}$) and the substantial rotation phase ($\Theta_g^{(1)}/\pi = -0.767$) are physically inseparable consequences of the same underlying mechanism: the rapid enhancement of continuum coupling and steep penetrability gradients ($dP_l/dE \gg 0$) immediately below the particle-emission threshold. The intense coupling to open continuum channels drives the pole deep into the complex energy plane (maximizing $\Gamma_g$), while strong interference with the non-resonant continuum and the neighboring poles twists the residue phase (driving $\cos2\Theta_g \to 0$ and shifting the strength almost entirely into the dispersive channel $\operatorname{Im}[\widetilde{B}_g]$). Consequently, the symmetric resonance peak is suppressed in the numerator and broadened by $(\Gamma_g/2)^2$ in the denominator, dissolving into a smooth background. 
Remarkably, despite its invisibility on the real axis, the intrinsic collectivity metrics of Pole~\#8 reach $N_{\text{tot}}\eta_g^{(1)} = 11.804$ and $C_g^{(1)2} N_{\text{tot}}\eta_g^{(1)} = 3.353$, which is fully comparable to that of the collective IVGDR state on $\text{RS}=5$ (Pole~\#1, $C_g^{(1)2} N_{\text{tot}}\eta_g^{(1)} = 3.492$). This demonstrates that extensive multi-channel configuration mixing is already fully established prior to channel opening, and underscores that real-axis strength functions alone can easily overlook strongly collective excitation modes residing deep in the complex energy plane.

Finally, the ratios $\sigma_g^{(c)} / \Gamma_g \approx 105\%\text{--}125\%$ in Table~\ref{tab:O16_E1_branching} reflect, through Eq.~(\ref{eq:R_first}), the energy variation of the rest of $\mathcal{S}_{cc}$ across the pole widths (Sec.~\ref{subsec:branching_ratio}).

\subsection{Isoscalar and isovector quadrupole ($L=2, \tau=0, 1$) continuum response}
\label{subsec:O16_L2_results}

We next extend the microscopic analysis to the quadrupole ($L=2$) continuum excitations in $^{16}\text{O}$ for both isoscalar ($\tau=0$) and isovector ($\tau=1$) transition operators. 
The total particle-hole channel dimension for $J^\pi = 2^+$ states is $N_{\text{tot}} = 16$ (comprising 8 neutron and 8 proton channels). 
The complex resonance poles identified across Riemann sheets 3 to 7 ($\text{RS}=3$--$7$), along with their open-channel Takagi singular values $\sigma_g^{(c)}$, collectivity indices, and complex transition strengths $\widetilde{B}(E2, \tau)/\pi$, are summarized in Table~\ref{tab:O16_E2_tau0_collectivity} for the isoscalar mode ($\tau=0$) and Table~\ref{tab:O16_E2_tau1_collectivity} for the isovector mode ($\tau=1$).

\begin{table*}[htbp]
\centering
\caption{Complex resonance pole energies $E_g = \operatorname{Re}(E_g) + i\operatorname{Im}(E_g)$, open-channel Takagi singular values $\sigma_g^{(c)}$, collectivity parameters ($C_g^{(0)2}, \Theta_g^{(0)}/\pi, N_{\text{tot}}\eta_g^{(0)}, C_g^{(0)2} N_{\text{tot}}\eta_g^{(0)}$ with $N_{\text{tot}} = 16$), and complex transition strengths $\operatorname{Re}[\widetilde{B}(E2, \tau=0)]/\pi$, $\operatorname{Im}[\widetilde{B}(E2, \tau=0)]/\pi$ evaluated via Eq.~(\ref{eq:B_contour}) for the isoscalar quadrupole ($L=2, \tau=0$) resonance poles in $^{16}\text{O}$ across Riemann sheets 3--7 ($\text{RS}=3$--$7$).}
\label{tab:O16_E2_tau0_collectivity}
\begin{ruledtabular}
\begin{tabular}{ccrrrrrrrrr}
\multirow{2}{*}{RS} & Pole & $\operatorname{Re}(E_g)$ & $\operatorname{Im}(E_g)$ & $\sigma_g^{(c)}$ & \multicolumn{4}{c}{Collectivity Parameters ($\tau=0$)} & \multicolumn{2}{c}{$\widetilde{B}(E2, \tau=0)/\pi$ [$e^2\text{fm}^4$]} \\
\cline{6-9} \cline{10-11}
 & No. & [MeV] & [MeV] & [MeV] & $C_g^{(0)2}$ & $\Theta_g^{(0)}/\pi$ & $N_{\text{tot}}\eta_g^{(0)}$ & $C_g^{(0)2} N_{\text{tot}}\eta_g^{(0)}$ & $\operatorname{Re}(\widetilde{B})/\pi$ & $\operatorname{Im}(\widetilde{B})/\pi$ \\
\hline
 3 & 1 & 16.767 & $-0.112$ & 0.233 & 0.9084 & $+0.992$ & 7.804 & 7.090 & $26.50$ & $-1.27$ \\
\hline
 \multirow{7}{*}{4} & 1 & 17.266 & $-1.642$ & 1.544 & 0.3140 & $-0.754$ & 5.990 & 1.881 & $6.65 \times 10^{-2}$ & $2.43$ \\
   & 2 & 18.747 & $-0.684$ & 0.047 & 0.0131 & $-0.969$ & 3.273 & 0.043 & $1.19 \times 10^{-4}$ & $2.38 \times 10^{-5}$ \\
   & 3 & 18.752 & $-0.714$ & 0.049 & 0.7692 & $-0.512$ & 3.164 & 2.434 & $-4.27 \times 10^{-3}$ & $3.29 \times 10^{-4}$ \\
   & 4 & 19.998 & $-1.274$ & 0.005 & 0.1546 & $-0.559$ & 7.891 & 1.220 & $-6.52 \times 10^{-5}$ & $2.56 \times 10^{-5}$ \\
   & 5 & 19.998 & $-1.272$ & 0.005 & 0.1575 & $+0.775$ & 5.954 & 0.938 & $5.23 \times 10^{-6}$ & $-3.25 \times 10^{-5}$ \\
   & 6 & 20.751 & $-1.224$ & 0.688 & 0.0217 & $-0.039$ & 3.049 & 0.066 & $9.54 \times 10^{-3}$ & $-2.37 \times 10^{-3}$ \\
   & 7 & 20.936 & $-1.407$ & 0.611 & 0.4243 & $-0.043$ & 2.453 & 1.041 & $8.13 \times 10^{-2}$ & $-2.28 \times 10^{-2}$ \\
\hline
 \multirow{4}{*}{5} & 1 & 21.995 & $-1.520$ & 1.457 & 0.0170 & $-0.173$ & 3.250 & 0.055 & $1.30 \times 10^{-2}$ & $-2.49 \times 10^{-2}$ \\
   & 2 & 27.028 & $-1.486$ & 3.759 & 0.0055 & $+0.323$ & 4.363 & 0.024 & $-4.41 \times 10^{-2}$ & $8.89 \times 10^{-2}$ \\
   & 3 & 28.503 & $-0.567$ & 1.245 & 0.0041 & $+0.617$ & 5.698 & 0.023 & $-2.82 \times 10^{-2}$ & $-2.55 \times 10^{-2}$ \\
   & 4 & 30.077 & $-0.658$ & 1.585 & 0.0007 & $+0.463$ & 7.046 & 0.005 & $-7.21 \times 10^{-3}$ & $1.68 \times 10^{-3}$ \\
\hline
 \multirow{3}{*}{6} & 1 & 33.635 & $-0.989$ & 0.072 & 0.0053 & $-0.926$ & 7.784 & 0.041 & $1.71 \times 10^{-5}$ & $8.58 \times 10^{-6}$ \\
   & 2 & 33.664 & $-0.914$ & 0.082 & 0.0278 & $-0.214$ & 5.195 & 0.144 & $1.61 \times 10^{-4}$ & $-7.07 \times 10^{-4}$ \\
   & 3 & 34.779 & $-0.591$ & 1.098 & 0.0286 & $-0.928$ & 5.361 & 0.153 & $6.98 \times 10^{-2}$ & $3.42 \times 10^{-2}$ \\
\hline
 \multirow{2}{*}{7} & 1 & 36.409 & $-1.899$ & 0.634 & 0.0098 & $-0.154$ & 8.331 & 0.082 & $4.66 \times 10^{-4}$ & $-6.79 \times 10^{-4}$ \\
   & 2 & 36.637 & $-0.423$ & 0.863 & 0.0130 & $+0.072$ & 5.176 & 0.067 & $2.73 \times 10^{-2}$ & $1.32 \times 10^{-2}$ \\
\end{tabular}
\end{ruledtabular}
\end{table*}

\begin{table*}[htbp]
\centering
\caption{Complex resonance pole energies $E_g = \operatorname{Re}(E_g) + i\operatorname{Im}(E_g)$, open-channel Takagi singular values $\sigma_g^{(c)}$, collectivity parameters ($C_g^{(1)2}, \Theta_g^{(1)}/\pi, N_{\text{tot}}\eta_g^{(1)}, C_g^{(1)2} N_{\text{tot}}\eta_g^{(1)}$ with $N_{\text{tot}} = 16$), and complex transition strengths $\operatorname{Re}[\widetilde{B}(E2, \tau=1)]/\pi$, $\operatorname{Im}[\widetilde{B}(E2, \tau=1)]/\pi$ evaluated via Eq.~(\ref{eq:B_contour}) for the isovector quadrupole ($L=2, \tau=1$) resonance poles in $^{16}\text{O}$ across Riemann sheets 3--7 ($\text{RS}=3$--$7$).}
\label{tab:O16_E2_tau1_collectivity}
\begin{ruledtabular}
\begin{tabular}{ccrrrrrrrrr}
\multirow{2}{*}{RS} & Pole & $\operatorname{Re}(E_g)$ & $\operatorname{Im}(E_g)$ & $\sigma_g^{(c)}$ & \multicolumn{4}{c}{Collectivity Parameters ($\tau=1$)} & \multicolumn{2}{c}{$\widetilde{B}(E2, \tau=1)/\pi$ [$e^2\text{fm}^4$]} \\
\cline{6-9} \cline{10-11}
 & No. & [MeV] & [MeV] & [MeV] & $C_g^{(1)2}$ & $\Theta_g^{(1)}/\pi$ & $N_{\text{tot}}\eta_g^{(1)}$ & $C_g^{(1)2} N_{\text{tot}}\eta_g^{(1)}$ & $\operatorname{Re}(\widetilde{B})/\pi$ & $\operatorname{Im}(\widetilde{B})/\pi$ \\
\hline
 3 & 1 & 16.767 & $-0.112$ & 0.233 & 0.0009 & $+0.308$ & 7.804 & 0.007 & $-9.13 \times 10^{-3}$ & $2.40 \times 10^{-2}$ \\
\hline
 \multirow{7}{*}{4} & 1 & 17.266 & $-1.642$ & 1.544 & 0.1107 & $-0.182$ & 5.990 & 0.663 & $3.56 \times 10^{-1}$ & $-7.78 \times 10^{-1}$ \\
   & 2 & 18.747 & $-0.684$ & 0.047 & 0.2366 & $-0.115$ & 3.273 & 0.775 & $1.65 \times 10^{-3}$ & $-1.46 \times 10^{-3}$ \\
   & 3 & 18.752 & $-0.714$ & 0.049 & 0.4723 & $+0.388$ & 3.164 & 1.494 & $-2.00 \times 10^{-3}$ & $1.70 \times 10^{-3}$ \\
   & 4 & 19.998 & $-1.274$ & 0.005 & 0.1825 & $-0.710$ & 7.891 & 1.440 & $-2.05 \times 10^{-5}$ & $8.01 \times 10^{-5}$ \\
   & 5 & 19.998 & $-1.272$ & 0.005 & 0.0305 & $+0.558$ & 5.954 & 0.182 & $-5.97 \times 10^{-6}$ & $-2.28 \times 10^{-6}$ \\
   & 6 & 20.751 & $-1.224$ & 0.688 & 0.2432 & $+0.911$ & 3.049 & 0.741 & $9.35 \times 10^{-2}$ & $-5.81 \times 10^{-2}$ \\
   & 7 & 20.936 & $-1.407$ & 0.611 & 0.6671 & $+0.886$ & 2.453 & 1.636 & $1.00 \times 10^{-1}$ & $-8.70 \times 10^{-2}$ \\
\hline
 \multirow{4}{*}{5} & 1 & 21.995 & $-1.520$ & 1.457 & 0.2617 & $-0.256$ & 3.250 & 0.851 & $-1.63 \times 10^{-2}$ & $-4.33 \times 10^{-1}$ \\
   & 2 & 27.028 & $-1.486$ & 3.759 & 0.1685 & $-0.092$ & 4.363 & 0.735 & $2.53$ & $-1.65$ \\
   & 3 & 28.503 & $-0.567$ & 1.245 & 0.0627 & $+0.781$ & 5.698 & 0.357 & $1.11 \times 10^{-1}$ & $-5.69 \times 10^{-1}$ \\
   & 4 & 30.077 & $-0.658$ & 1.585 & 0.1562 & $-0.240$ & 7.046 & 1.100 & $9.92 \times 10^{-2}$ & $-1.55$ \\
\hline
 \multirow{3}{*}{6} & 1 & 33.635 & $-0.989$ & 0.072 & 0.1194 & $-0.901$ & 7.784 & 0.929 & $3.47 \times 10^{-4}$ & $2.50 \times 10^{-4}$ \\
   & 2 & 33.664 & $-0.914$ & 0.082 & 0.2124 & $+0.519$ & 5.195 & 1.103 & $-5.50 \times 10^{-3}$ & $-6.69 \times 10^{-4}$ \\
   & 3 & 34.779 & $-0.591$ & 1.098 & 0.1684 & $-0.188$ & 5.361 & 0.903 & $1.75 \times 10^{-1}$ & $-4.24 \times 10^{-1}$ \\
\hline
 \multirow{2}{*}{7} & 1 & 36.409 & $-1.899$ & 0.634 & 0.1446 & $-0.267$ & 8.331 & 1.205 & $-1.33 \times 10^{-3}$ & $-1.20 \times 10^{-2}$ \\
   & 2 & 36.637 & $-0.423$ & 0.863 & 0.3510 & $-0.222$ & 5.176 & 1.817 & $1.41 \times 10^{-1}$ & $-8.04 \times 10^{-1}$ \\
\end{tabular}
\end{ruledtabular}
\end{table*}

\begin{figure}[htbp]
\centering
\includegraphics[width=\columnwidth]{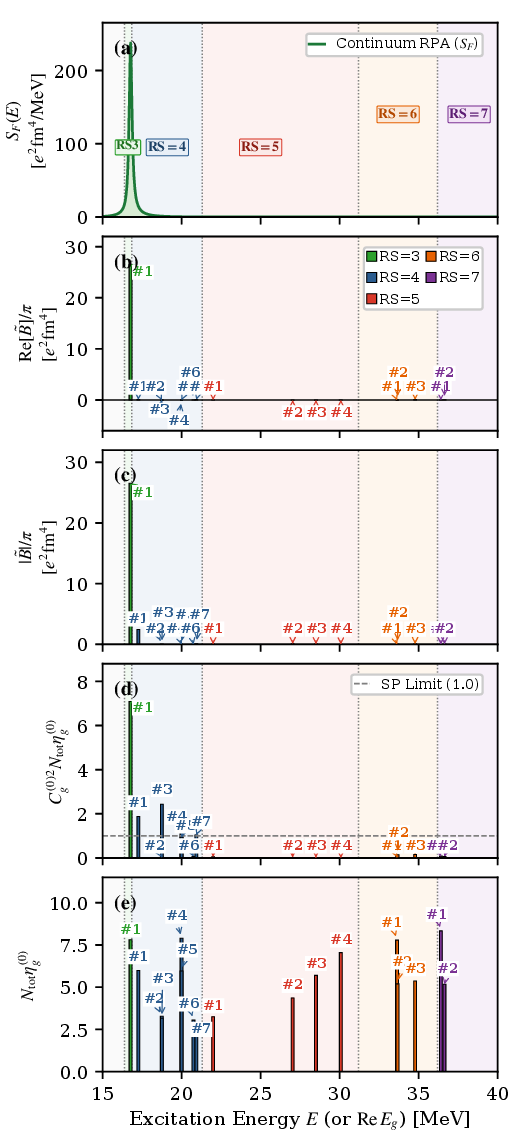}
\caption{(Color online) Same as Fig.~\ref{fig:O16_E1_analysis} but for the isoscalar quadrupole ($E2, \tau=0$) continuum response in $^{16}\text{O}$ across Riemann sheets 3 to 7 ($\text{RS}=3$--$7$) on full scale. The prominent peak on $\text{RS}=3$ corresponds to the isoscalar giant quadrupole resonance (ISGQR). See text for details.}
\label{fig:O16_E2_IS_full}
\end{figure}

\begin{figure}[htbp]
\centering
\includegraphics[width=\columnwidth]{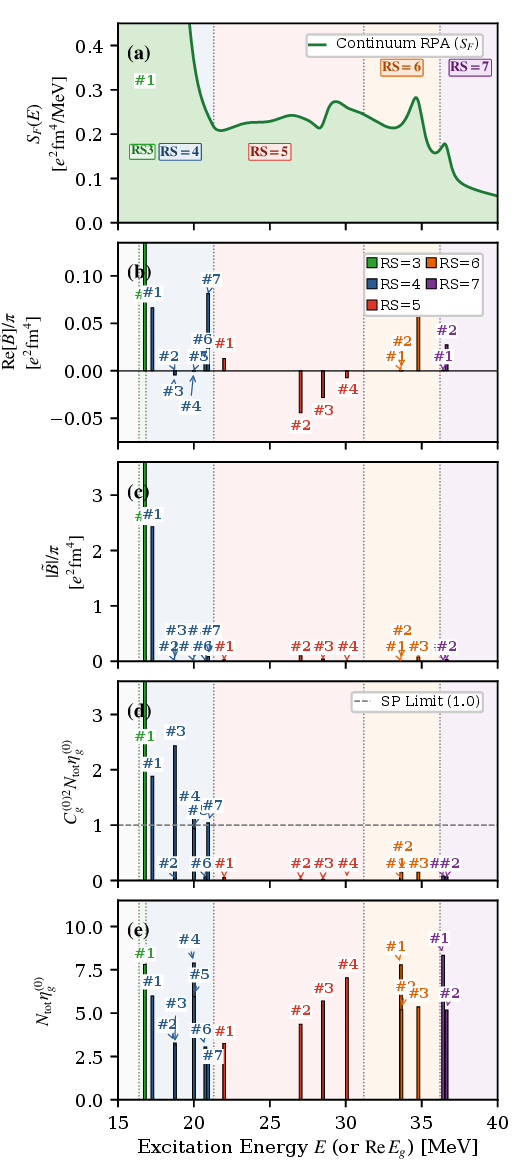}
\caption{(Color online) Same as Fig.~\ref{fig:O16_E2_IS_full} but with rescaled vertical axes in panels (a)--(d) to highlight the microscopic fine structures across $\text{RS}=4$--$7$. Panel (a) is scaled to $[0.0, 0.45]\ e^2\text{fm}^4/\text{MeV}$ matching Ref.~\cite{Mizuyama2025_JRPA2}. See text for details.}
\label{fig:O16_E2_IS_zoomed}
\end{figure}

\begin{figure}[htbp]
\centering
\includegraphics[width=\columnwidth]{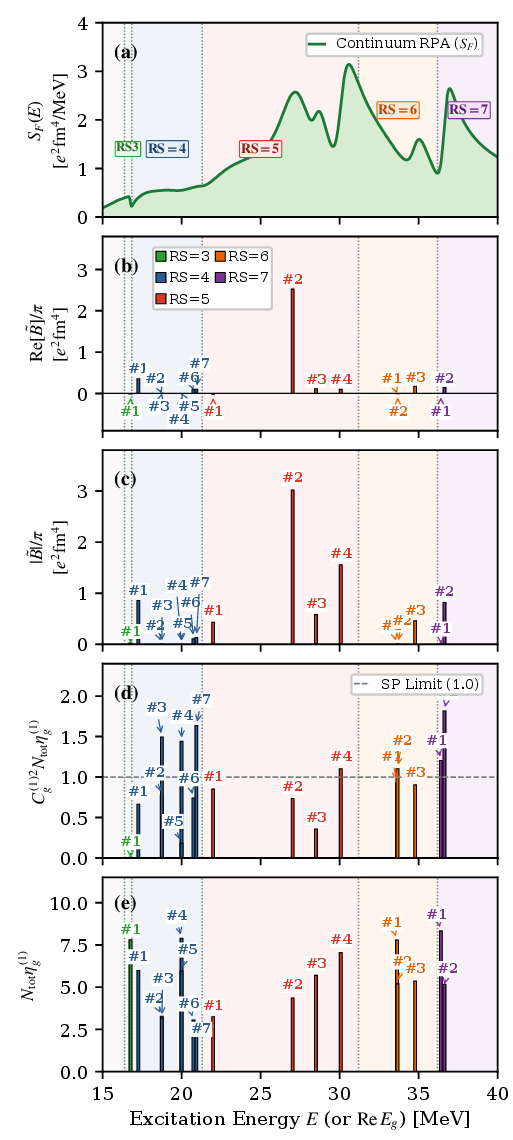}
\caption{(Color online) Same as Fig.~\ref{fig:O16_E1_analysis} but for the isovector quadrupole ($E2, \tau=1$) continuum response in $^{16}\text{O}$ across Riemann sheets 3 to 7 ($\text{RS}=3$--$7$). Shaded background regions denote the respective Riemann sheets separated by particle decay thresholds (vertical dotted lines). See text for details.}
\label{fig:O16_E2_IV_analysis}
\end{figure}

For the isoscalar quadrupole mode ($\tau=0$, Table~\ref{tab:O16_E2_tau0_collectivity}), the microscopic decomposition across Riemann sheets 3 to 7 is displayed in Figs.~\ref{fig:O16_E2_IS_full} and \ref{fig:O16_E2_IS_zoomed}. A prominent feature is the clean identification of the isoscalar giant quadrupole resonance (ISGQR) located on Riemann sheet 3 at $E_g = 16.767 - 0.112i\text{ MeV}$ (Pole~\#1). As shown in the full-scale decomposition [Fig.~\ref{fig:O16_E2_IS_full}], this state possesses an immense real transition strength $\operatorname{Re}[\widetilde{B}(E2, \tau=0)]/\pi = 26.50\ e^2\text{fm}^4$ and an exceptional collectivity enhancement factor $C_g^{(0)2} N_{\text{tot}}\eta_g^{(0)} = 7.090 \gg 1$, resulting from the in-phase coherent superposition of multiple proton and neutron particle-hole excitations ($C_g^{(0)2} = 0.9084$, $N_{\text{tot}}\eta_g^{(0)} = 7.804$, and $\Theta_g^{(0)}/\pi = +0.992$). In stark contrast, its isovector collectivity (Table~\ref{tab:O16_E2_tau1_collectivity}) at the same pole energy is quenched to $C_g^{(1)2} N_{\text{tot}}\eta_g^{(1)} = 0.007$, unequivocally demonstrating its pristine isoscalar giant character.

To inspect the fine structures obscured by the dominant ISGQR peak, Fig.~\ref{fig:O16_E2_IS_zoomed} displays the rescaled decomposition, where panel (a) is scaled to $[0.0, 0.45]\ e^2\text{fm}^4/\text{MeV}$ in line with Ref.~\cite{Mizuyama2025_JRPA2}. In this high-energy continuum region ($27\text{--}30\text{ MeV}$ on $\text{RS}=5$), Poles~\#2, \#3, and \#4 exhibit negative transition residues $\operatorname{Re}[\widetilde{B}(E2, \tau=0)]/\pi < 0$ [Fig.~\ref{fig:O16_E2_IS_zoomed}(b)], producing characteristic dips in the continuum RPA strength function $S_F(E)$ [Fig.~\ref{fig:O16_E2_IS_zoomed}(a)]. Although their collectivity factors are heavily quenched ($C_g^{(0)2} N_{\text{tot}}\eta_g^{(0)} \ll 1.0$) due to phase cancellation between protons and neutrons in the isoscalar channel, their raw channel coherence sums remain substantial ($N_{\text{tot}}\eta_g^{(0)} \approx 4.4\text{--}7.0$, Fig.~\ref{fig:O16_E2_IS_zoomed}(e)), reflecting strong multi-channel coupling to the continuum.

\begin{table*}[htbp]
\centering
\caption{Partial decay widths $\Gamma_i = \sigma_g^{(c)} |\omega_{g,i}^{(c)}|^2$ and relative branching ratios $\mathrm{br}_i = |\omega_{g,i}^{(c)}|^2$ evaluated via rank-1 Takagi factorization for the two key isoscalar quadrupole ($L=2, \tau=0$) resonance poles in $^{16}\text{O}$: the ISGQR peak on $\text{RS}=3$ (Pole~\#1, $16.77\text{ MeV}$) and the pole on $\text{RS}=5$ at $27.03\text{ MeV}$ (Pole~\#2), whose pole term is of dip type ($\operatorname{Re}\widetilde{B}<0$, $q=-0.62$).}
\label{tab:O16_E2_tau0_branching}
\begin{ruledtabular}
\begin{tabular}{ccrrrr}
\makecell[c]{Nucleon\\Type} & \makecell[c]{Decay Channel\\(Hole $\to$ Particle Wave)} & \multicolumn{2}{c}{RS=3, Pole \#1 ($16.77\text{ MeV}$)} & \multicolumn{2}{c}{RS=5, Pole \#2 ($27.03\text{ MeV}$)} \\
\cline{3-4} \cline{5-6}
 & & $\Gamma_i$ [MeV] & $\mathrm{br}_i$ [\%] & $\Gamma_i$ [MeV] & $\mathrm{br}_i$ [\%] \\
\hline
\multirow{6}{*}{Neutron} 
 & $1p_{1/2} \to p_{3/2}\ (\Delta l = 0)$  & 0.0406 & 17.42 & 0.0107 &  0.28 \\
 & $1p_{1/2} \to f_{5/2}\ (\Delta l = +2)$ & $1.69 \times 10^{-5}$ &  0.01 & 0.0726 &  1.93 \\
 & $1p_{3/2} \to f_{5/2}\ (\Delta l = +2)$ & \multicolumn{2}{c}{---} & 0.0014 &  0.04 \\
 & $1p_{3/2} \to f_{7/2}\ (\Delta l = +2)$ & \multicolumn{2}{c}{---} & 0.6103 & 16.24 \\
 & $1p_{3/2} \to p_{1/2}\ (\Delta l = 0)$  & \multicolumn{2}{c}{---} & 0.0064 &  0.17 \\
 & $1p_{3/2} \to p_{3/2}\ (\Delta l = 0)$  & \multicolumn{2}{c}{---} & 0.0099 &  0.26 \\
\hline
\multirow{6}{*}{Proton} 
 & $1p_{1/2} \to p_{3/2}\ (\Delta l = 0)$  & 0.1725 & 74.07 & 0.0414 &  1.10 \\
 & $1p_{1/2} \to f_{5/2}\ (\Delta l = +2)$ & 0.0198 &  8.50 & 0.5400 & 14.37 \\
 & $1p_{3/2} \to f_{5/2}\ (\Delta l = +2)$ & \multicolumn{2}{c}{---} & 0.0282 &  0.75 \\
 & $1p_{3/2} \to f_{7/2}\ (\Delta l = +2)$ & \multicolumn{2}{c}{---} & 2.3965 & 63.75 \\
 & $1p_{3/2} \to p_{1/2}\ (\Delta l = 0)$  & \multicolumn{2}{c}{---} & 0.0112 &  0.30 \\
 & $1p_{3/2} \to p_{3/2}\ (\Delta l = 0)$  & \multicolumn{2}{c}{---} & 0.0304 &  0.81 \\
\hline
\multicolumn{2}{l}{Total Open Width $\sigma_g^{(c)}$ [MeV]} & \multicolumn{2}{c}{0.2329} & \multicolumn{2}{c}{3.7589} \\
\multicolumn{2}{l}{Total Pole Width $\Gamma_g$ [MeV]} & \multicolumn{2}{c}{0.2242} & \multicolumn{2}{c}{2.9714} \\
\multicolumn{2}{l}{Ratio $\sigma_g^{(c)} / \Gamma_g$} & \multicolumn{2}{c}{103.9\%} & \multicolumn{2}{c}{126.5\%} \\
\multicolumn{2}{l}{Neutron / Proton Sum [\%]} & \multicolumn{2}{c}{17.4\% / 82.6\%} & \multicolumn{2}{c}{18.9\% / 81.1\%} \\
\end{tabular}
\end{ruledtabular}
\end{table*}

To gain microscopic insight into the decay dynamics of the ISGQR and the antiresonant dip states, Table~\ref{tab:O16_E2_tau0_branching} lists the partial decay widths $\Gamma_i$ and relative branching ratios $\mathrm{br}_i$ evaluated via rank-1 Takagi factorization. For the ISGQR peak (RS=3, Pole~\#1 at $16.77\text{ MeV}$), the deeper $1p_{3/2}^{-1}$ hole channels remain energetically closed, so that decay proceeds exclusively ($100.0\%$) via the valence $1p_{1/2}^{-1}$ ground-state transitions ($p(1p_{1/2} \to p_{3/2})$ carrying $74.07\%$ and $n(1p_{1/2} \to p_{3/2})$ carrying $17.42\%$, with proton decay favored by the lower Coulomb threshold).
In stark contrast, for Pole~\#2 on $\text{RS}=5$ ($27.03\text{ MeV}$), whose pole term is of dip type, a dramatic reversal of decay dominance occurs: 
as detailed in Table~\ref{tab:O16_E2_tau0_branching}, the decay is overwhelmingly dominated by the deep inner-shell $1p_{3/2}^{-1}$ hole configurations, which account for a combined branching ratio of $82.32\%$ (obtained by summing all eight $1p_{3/2}^{-1}$ decay channels: proton $1p_{3/2}^{-1}$ contributing $65.61\%$, which is heavily concentrated in $p(1p_{3/2} \to f_{7/2})$ at $63.75\%$, and neutron $1p_{3/2}^{-1}$ contributing $16.71\%$, concentrated in $n(1p_{3/2} \to f_{7/2})$ at $16.24\%$, so that the $1p_{3/2}^{-1} \to f_{7/2}$ transitions alone constitute $79.99\%$). In contrast, the ground-state valence decay channels ($1p_{1/2}^{-1}$) drop to a minor fraction of only $17.68\%$ ($15.47\%$ proton plus $2.21\%$ neutron).
 Because the $1p_{3/2}^{-1}$ channels are open on $\text{RS}=5$, this pole decays mainly through the channels of its own dominant configurations. The branching ratios, which carry no phase information, therefore characterize its decay but do not by themselves establish a quasi-bound state embedded in the valence continuum, nor the interference that shapes $S_F(E)$; the latter is governed by the phase $\Theta_g$ through Eq.~(\ref{eq:fano_q_theta_mapping}).

To establish the microscopic origin of the Fano asymmetry parameter $q$ and its algebraic connection with the present Jost RPA framework, we examine Fano's configuration interaction definition \cite{Fano1961}:
\begin{equation}
q = \frac{\langle \Phi_{\text{disc}} | \hat{T} | \Psi_0 \rangle + \mathcal{P} \int dE' \frac{V_{E'} \langle \psi_{E'} | \hat{T} | \Psi_0 \rangle}{E - E'}}{\pi V_E^* \langle \psi_E | \hat{T} | \Psi_0 \rangle},
\label{eq:fano_q_definition}
\end{equation}
where $|\Psi_0\rangle$ is the nuclear ground state, $|\Phi_{\text{disc}}\rangle$ denotes the discrete quasi-bound state modified by continuum coupling, $|\psi_E\rangle$ represents the unperturbed continuum channel state, $V_E = \langle \psi_E | \hat{H} | \Phi_{\text{disc}}\rangle$ is the configuration-interaction coupling matrix element, and $\hat{T}$ is the excitation operator (here the quadrupole operator $\hat{Q}_{20}$).
In Fano's formalism, the transition cross section in the presence of an autoionizing or quasi-bound state embedded in a continuous spectrum is expressed as
\begin{equation}
\sigma(E) = \sigma_0 \frac{(q + \epsilon)^2}{1 + \epsilon^2},
\label{eq:fano_formula}
\end{equation}
where $\epsilon = (E - \operatorname{Re}E_g)/(\Gamma_g/2)$ is the dimensionless reduced energy, and $\sigma_0$ denotes the non-resonant background continuum cross section. By expanding the numerator as $(q + \epsilon)^2 = (q^2 - 1) + 2q\epsilon + (1 + \epsilon^2)$, Eq.~(\ref{eq:fano_formula}) is decomposed into a non-resonant background plus symmetric and asymmetric resonant contributions:
\begin{equation}
\sigma(E) = \sigma_0 \left[ 1 + \frac{q^2 - 1}{1 + \epsilon^2} + \frac{2q\epsilon}{1 + \epsilon^2} \right].
\label{eq:fano_expanded}
\end{equation}

On the other hand, in the vicinity of an isolated complex resonance pole $E_g$, the continuum RPA strength function $S_F(E)$ in the Jost RPA Mittag-Leffler expansion is written with the complex transition residue $\widetilde{B}_g = |\widetilde{B}_g| e^{2i\Theta_g}$ ($\Theta_g=\arg\mathcal{F}_g^{(\tau)}$, the phase listed in Tables~\ref{tab:O16_E2_tau0_collectivity} and \ref{tab:O16_E2_tau1_collectivity}) as
\begin{align}
S_F(E) &= S_{\text{bg}}(E) {-} \frac{1}{\pi} \operatorname{Im}\left[ \frac{\widetilde{B}_g}{E - E_g} \right] \nonumber \\
&{= S_{\text{bg}}(E) + \frac{|\widetilde{B}_g|}{\pi (\Gamma_g/2)} \left[ \frac{\cos2\Theta_g}{1 + \epsilon^2} - \frac{\sin2\Theta_g\, \epsilon}{1 + \epsilon^2} \right],}
\label{eq:jost_mittag_leffler}
\end{align}
where $S_{\text{bg}}(E)$ is the smooth non-resonant continuum background. Direct term-by-term comparison between the resonant profile in Eq.~(\ref{eq:fano_expanded}) and Eq.~(\ref{eq:jost_mittag_leffler}) shows that under the approximation of an isolated resonance pole embedded in a locally smooth continuum background, the ratio of the symmetric Lorentzian component to the asymmetric dispersive component satisfies
\begin{equation}
\frac{q^2 - 1}{2q} = \frac{\cos2\Theta_g}{-\sin2\Theta_g} = -\cot2\Theta_g.
\label{eq:fano_phase_matching}
\end{equation}
 The quadratic equation $q^2 + 2q\cot2\Theta_g - 1 = 0$ admits two algebraic branches, $q = -\cot2\Theta_g \pm \csc2\Theta_g$, which evaluate via trigonometric half-angle identities to $q = -\cot\Theta_g$ and $q = \tan\Theta_g$. Matching the dispersive terms gives $\sigma_0=-|\widetilde{B}_g|\sin2\Theta_g/[2q\,\pi(\Gamma_g/2)]$, which is non-negative only for $q=-\cot\Theta_g$, where $\sigma_0=|\widetilde{B}_g|\sin^2\Theta_g/[\pi(\Gamma_g/2)]$. We therefore select the physical branch
\begin{equation}
q = -\cot\Theta_g,
\label{eq:fano_q_theta_mapping}
\end{equation}
 which establishes the algebraic representation of the Fano asymmetry parameter in terms of the phase of the transition amplitude. The antiresonance dip limit ($q \to 0$) corresponds to $\Theta_g \to \pm\pi/2$, i.e., $\widetilde{B}_g\to-|\widetilde{B}_g|$, while the Breit-Wigner peak limit ($|q| \to \infty$) corresponds to $\Theta_g \to 0$ or $\pi$, i.e., $\widetilde{B}_g\to+|\widetilde{B}_g|$. For the ISGQR ($\Theta_g^{(0)}/\pi=+0.992$) this gives $q=+39.8$, a symmetric Breit-Wigner peak.

This effective algebraic formulation provides a transparent microscopic interpretation for the emergence of high-energy Fano-like dips in the isoscalar quadrupole spectrum:
\begin{enumerate}[(i)]
\item \emph{Concentration of the low-lying ISGQR strength:} Because the isoscalar particle-hole residual interaction $V_{\text{res}}^{\tau=0}$ is strongly attractive, it constructively adds the transition matrix elements of all unperturbed $2\hbar\omega$ particle-hole pairs in-phase and shifts the collective ISGQR down to $E_g = 16.767 - 0.112i\text{ MeV}$ on $\text{RS}=3$. This single giant resonance strongly concentrates the low-energy quadrupole transition strength ($C_g^{(0)2} N_{\text{tot}}\eta_g^{(0)} = 7.090 \gg 1$).
\item \emph{Direct transition depletion of high-energy core excitations ($T_{\text{direct}} \to 0$):} By state orthogonality and the concentration of low-energy transition strength into the ISGQR, the remaining particle-hole states located in the high-energy continuum region ($27\text{--}30\text{ MeV}$) are left with virtually zero direct transition matrix element from the ground state ($\langle \Phi_{\text{disc}} | \hat{Q}_{20} | \Psi_0 \rangle \approx 0$). As discussed above, their decay is dominated by the deep inner-shell $1p_{3/2}^{-1}$ hole configurations (Table~\ref{tab:O16_E2_tau0_branching}).
\item \emph{Destructive antiresonance ($q \to 0$, $\Theta_g \to \pm\pi/2$):} Because the direct excitation amplitude in the numerator of Eq.~(\ref{eq:fano_q_definition}) is quenched ($\langle \Phi_{\text{disc}} | \hat{Q}_{20} | \Psi_0 \rangle \to 0$), the discrete state can only couple to the continuum in anti-phase via the residual interaction $V_E$, which cancels the background continuum excitation. This drives the numerator to zero, compelling $q \to 0$, i.e., $\Theta_g \to \pm\pi/2$ [Eq.~(\ref{eq:fano_q_theta_mapping})]. The cross section in Eq.~(\ref{eq:fano_formula}) then reduces to:
\begin{equation}
\sigma(E) = \sigma_0 \frac{\epsilon^2}{1 + \epsilon^2} = \sigma_0 \left( 1 - \frac{1}{1 + \epsilon^2} \right),
\label{eq:fano_dip_shape}
\end{equation}
which vanishes identically ($\sigma(\operatorname{Re}E_g) = 0$) at the resonance energy $\epsilon = 0$. In general the single-pole profile vanishes at $\epsilon=-q$, i.e., at $E=\operatorname{Re}E_g-q\Gamma_g/2$. For the $\text{RS}=5$ poles of Table~\ref{tab:O16_E2_tau0_collectivity}, $q=-\cot\Theta_g^{(0)}$ gives $q=-0.62$ for Pole~\#2 (zero at $27.95\text{ MeV}$), $q=+0.385$ for Pole~\#3 (zero at $28.29\text{ MeV}$) and $q\simeq-0.12$ for Pole~\#4 (zero at $30.15\text{ MeV}$). Pole~\#4 is thus the closest to the pure antiresonance limit, and the minimum of $S_F(E)$ observed at $28.25\text{ MeV}$ is associated with Pole~\#3: removing its pole term from $S_F(E)$ removes the minimum. 
\end{enumerate}
This suggests that the intense strength concentration in the low-lying ISGQR and the emergence of pure Fano antiresonance dips at higher excitation energies are inseparable, mutually complementary manifestations of nuclear continuum dynamics.

For the isovector quadrupole response ($\tau=1$), the microscopic breakdown across Riemann sheets 3 to 7 is displayed in Fig.~\ref{fig:O16_E2_IV_analysis} and summarized in Table~\ref{tab:O16_E2_tau1_collectivity}. The shaded background regions in Figs.~\ref{fig:O16_E2_IS_full}--\ref{fig:O16_E2_IV_analysis} distinguish $\text{RS}=2$ ($E < 16.39\text{ MeV}$), $\text{RS}=3$ ($16.39 \le E < 16.84\text{ MeV}$), $\text{RS}=4$ ($16.84 \le E < 21.31\text{ MeV}$), $\text{RS}=5$ ($21.31 \le E < 31.19\text{ MeV}$), $\text{RS}=6$ ($31.19 \le E < 36.20\text{ MeV}$), and $\text{RS}=7$ ($E \ge 36.20\text{ MeV}$), separated by the particle decay thresholds (vertical dotted lines). 
A crucial insight provided by the present Takagi decomposition concerns Pole~\#2 on $\text{RS}=5$ ($E_g = 27.028 - 1.486i\text{ MeV}$). As shown in Figs.~\ref{fig:O16_E2_IV_analysis}(b) and (c), this state produces the most prominent peak in the transition strength with $\operatorname{Re}[\widetilde{B}(E2, \tau=1)]/\pi = 2.53\ e^2\text{fm}^4$ and $|\widetilde{B}(E2, \tau=1)|/\pi = 3.02\ e^2\text{fm}^4$. 
However, its collectivity factor [Fig.~\ref{fig:O16_E2_IV_analysis}(d)] is only $C_g^{(1)2} N_{\text{tot}}\eta_g^{(1)} = 0.735$, remaining below the single-particle limit ($=1.0$). This demonstrates that the large $B(E2)$ value of Pole~\#2 does not originate from coherent multi-channel configuration mixing, but is predominantly driven by a single unperturbed particle-hole continuum transition possessing a large spatial overlap $\langle r^2 Y_2 \rangle$ with extended continuum wave functions (a quasi-single-particle continuum resonance).

Instead, the collective components of the isovector giant quadrupole resonance (IVGQR) identified by the present metric ($C_g^{(1)2} N_{\text{tot}}\eta_g^{(1)} > 1.0$) are pushed up to higher excitation energies and fragmented across multiple Riemann sheets due to the repulsive isovector residual interaction and strong continuum coupling. Specifically, Pole~\#4 on $\text{RS}=5$ ($E_g = 30.077 - 0.658i\text{ MeV}$) and Pole~\#2 on $\text{RS}=7$ ($E_g = 36.637 - 0.423i\text{ MeV}$) exhibit enhanced collectivity indices of $C_g^{(1)2} N_{\text{tot}}\eta_g^{(1)} = 1.100$ and $1.817$, respectively. 
This microscopic decomposition provides a clear and plausible physical scenario for understanding the long-standing controversy regarding the structure of the IVGQR in $^{16}\text{O}$, explaining within the present continuum RPA framework why experimental searches \cite{Hanna1974, Snover1974, Phillips1979} and earlier continuum RPA calculations \cite{Krewald1974, Speth1981} observed broad, fragmented E2 strengths distributed over $25\text{--}40\text{ MeV}$ rather than a single compact collective peak.

\begin{table*}[htbp]
\centering
\caption{Partial decay widths $\Gamma_i = \sigma_g^{(c)} |\omega_{g,i}^{(c)}|^2$ and relative branching ratios $\mathrm{br}_i = |\omega_{g,i}^{(c)}|^2$ evaluated via rank-1 Takagi factorization for the three key isovector quadrupole ($L=2, \tau=1$) resonance poles in $^{16}\text{O}$: the quasi-single-particle resonance on $\text{RS}=5$ (Pole~\#2, $27.03\text{ MeV}$), the low-energy collective IVGQR fragment on $\text{RS}=5$ (Pole~\#4, $30.08\text{ MeV}$), and the global maximum collective IVGQR state on $\text{RS}=7$ (Pole~\#2, $36.64\text{ MeV}$).}
\label{tab:O16_E2_tau1_branching}
\begin{ruledtabular}
\begin{tabular}{ccrrrrrr}
\makecell[c]{Nucleon\\Type} & \makecell[c]{Decay Channel\\(Hole $\to$ Particle Wave)} & \multicolumn{2}{c}{RS=5, Pole \#2 ($27.03\text{ MeV}$)} & \multicolumn{2}{c}{RS=5, Pole \#4 ($30.08\text{ MeV}$)} & \multicolumn{2}{c}{RS=7, Pole \#2 ($36.64\text{ MeV}$)} \\
\cline{3-4} \cline{5-6} \cline{7-8}
 & & $\Gamma_i$ [MeV] & $\mathrm{br}_i$ [\%] & $\Gamma_i$ [MeV] & $\mathrm{br}_i$ [\%] & $\Gamma_i$ [MeV] & $\mathrm{br}_i$ [\%] \\
\hline
\multirow{8}{*}{Neutron} 
 & $1p_{1/2} \to p_{3/2}\ (\Delta l = 0)$  & 0.0107 &  0.28 & 0.0329 &  2.08 & 0.0240 &  2.78 \\
 & $1p_{1/2} \to f_{5/2}\ (\Delta l = +2)$ & 0.0726 &  1.93 & 0.2317 & 14.62 & 0.1573 & 18.22 \\
 & $1p_{3/2} \to f_{5/2}\ (\Delta l = +2)$ & 0.0014 &  0.04 & 0.0165 &  1.04 & 0.0513 &  5.94 \\
 & $1p_{3/2} \to f_{7/2}\ (\Delta l = +2)$ & 0.6103 & 16.24 & 0.3966 & 25.02 & 0.1565 & 18.12 \\
 & $1p_{3/2} \to p_{1/2}\ (\Delta l = 0)$  & 0.0064 &  0.17 & 0.0421 &  2.66 & 0.0217 &  2.51 \\
 & $1p_{3/2} \to p_{3/2}\ (\Delta l = 0)$  & 0.0099 &  0.26 & 0.0394 &  2.49 & 0.0229 &  2.65 \\
 & $1s_{1/2} \to d_{3/2}\ (\Delta l = +2)$ & \multicolumn{2}{c}{---} & \multicolumn{2}{c}{---} & 0.0384 &  4.45 \\
 & $1s_{1/2} \to d_{5/2}\ (\Delta l = +2)$ & \multicolumn{2}{c}{---} & \multicolumn{2}{c}{---} & 0.0007 &  0.09 \\
\hline
\multirow{8}{*}{Proton} 
 & $1p_{1/2} \to p_{3/2}\ (\Delta l = 0)$  & 0.0414 &  1.10 & 0.0004 &  0.03 & 0.0083 &  0.96 \\
 & $1p_{1/2} \to f_{5/2}\ (\Delta l = +2)$ & 0.5400 & 14.37 & 0.2524 & 15.93 & 0.1132 & 13.11 \\
 & $1p_{3/2} \to f_{5/2}\ (\Delta l = +2)$ & 0.0282 &  0.75 & 0.0294 &  1.86 & 0.0355 &  4.11 \\
 & $1p_{3/2} \to f_{7/2}\ (\Delta l = +2)$ & 2.3965 & 63.75 & 0.5366 & 33.85 & 0.1353 & 15.67 \\
 & $1p_{3/2} \to p_{1/2}\ (\Delta l = 0)$  & 0.0112 &  0.30 & 0.0060 &  0.38 & 0.0022 &  0.25 \\
 & $1p_{3/2} \to p_{3/2}\ (\Delta l = 0)$  & 0.0304 &  0.81 & 0.0010 &  0.06 & 0.0047 &  0.54 \\
 & $1s_{1/2} \to d_{3/2}\ (\Delta l = +2)$ & \multicolumn{2}{c}{---} & \multicolumn{2}{c}{---} & 0.0767 &  8.89 \\
 & $1s_{1/2} \to d_{5/2}\ (\Delta l = +2)$ & \multicolumn{2}{c}{---} & \multicolumn{2}{c}{---} & 0.0149 &  1.72 \\
\hline
\multicolumn{2}{l}{Total Open Width $\sigma_g^{(c)}$ [MeV]} & \multicolumn{2}{c}{3.7589} & \multicolumn{2}{c}{1.5851} & \multicolumn{2}{c}{0.8633} \\
\multicolumn{2}{l}{Total Pole Width $\Gamma_g$ [MeV]} & \multicolumn{2}{c}{2.9714} & \multicolumn{2}{c}{1.3157} & \multicolumn{2}{c}{0.8455} \\
\multicolumn{2}{l}{Ratio $\sigma_g^{(c)} / \Gamma_g$} & \multicolumn{2}{c}{126.5\%} & \multicolumn{2}{c}{120.5\%} & \multicolumn{2}{c}{102.1\%} \\
\multicolumn{2}{l}{Neutron / Proton Sum [\%]} & \multicolumn{2}{c}{18.9\% / 81.1\%} & \multicolumn{2}{c}{47.9\% / 52.1\%} & \multicolumn{2}{c}{50.7\% / 49.3\%} \\
\end{tabular}
\end{ruledtabular}
\end{table*}

To substantiate the physical distinction between the quasi-single-particle resonance and the collective IVGQR components identified by the present diagnostic, Table~\ref{tab:O16_E2_tau1_branching} compares the partial decay widths $\Gamma_i$ and relative branching ratios $\mathrm{br}_i$ for these three key poles. 
For Pole~\#2 on $\text{RS}=5$ ($27.03\text{ MeV}$), the decay is heavily concentrated into a single channel, with the proton $p(1p_{3/2} \to f_{7/2})$ transition accounting for $63.75\%$ of the total width (total proton decay $= 81.1\%$). This pronounced single-channel concentration confirms its quasi-single-particle continuum character.
In sharp contrast, for Pole~\#4 on $\text{RS}=5$ ($30.08\text{ MeV}$), the decay width is evenly shared between neutrons ($47.9\%$) and protons ($52.1\%$) and dispersed across multiple $1p_{3/2}$ and $1p_{1/2}$ channels ($p(1p_{3/2} \to f_{7/2}): 33.85\%$, $n(1p_{3/2} \to f_{7/2}): 25.02\%$, $p(1p_{1/2} \to f_{5/2}): 15.93\%$, and $n(1p_{1/2} \to f_{5/2}): 14.62\%$).
Furthermore, for the highest-collectivity IVGQR state on $\text{RS}=7$ (Pole~\#2 at $36.64\text{ MeV}$, $C_g^{(1)2} N_{\text{tot}}\eta_g^{(1)} = 1.817$), all 16 channels actively participate in the decay, including the deepest $1s_{1/2}^{-1}$ core-hole transitions ($15.15\%$). The total decay exhibits near-perfect isospin symmetry (neutrons $50.7\%$ : protons $49.3\%$), providing a clear microscopic demonstration of coherent multi-channel collectivity in the deep continuum.

It is instructive to discuss the qualitative implications of these decay branching patterns in connection with experimental radiative capture studies, such as $^{15}\text{N}(p, \gamma_0)^{16}\text{O}$ and polarized $(\vec{p}, \gamma)$ reactions \cite{Hanna1974, Snover1974, Phillips1979}. In those experiments, prominent $E2$ strength was observed in the $25\text{--}30\text{ MeV}$ excitation region, with strong selectivity toward proton channels. The present finding that the $27.03\text{ MeV}$ quasi-single-particle resonance is overwhelmingly dominated by proton decay ($81.1\%$ total proton decay, with $63.75\%$ into the $p(1p_{3/2} \to f_{7/2})$ channel) provides a plausible microscopic scenario explaining why proton-induced capture experiments are particularly sensitive to this continuum structure. Nevertheless, these numerical results should be viewed as pointing out a physical possibility within the present model, whose limitations are summarized in Sec.~\ref{sec:summary}.

Finally, we elucidate the physical mechanism behind the prominent $f_{7/2}$ single-particle strength and the behavior of the ratio $\sigma_g^{(c)} / \Gamma_g$ in Tables~\ref{tab:O16_E2_tau0_branching} and \ref{tab:O16_E2_tau1_branching}. For electric quadrupole excitations ($L=2$), transitions from $p$-shell holes ($l=1$) into $f$-wave continuum states ($l=3$) encounter a formidable centrifugal barrier $V_{\text{cent}}(r) = \hbar^2 l(l+1)/(2mr^2)$ with $l(l+1) = 12$, which is six times steeper than that for $p$-wave continuum states ($l(l+1) = 2$). This large barrier effectively traps low-energy continuum particles inside the nuclear interior, forming a core-excited quasi-bound resonance with large radial transition matrix elements $\langle r^2 Y_2 \rangle$ and producing the prominent single-particle peak at $27.03\text{ MeV}$.
 The ratio $\sigma_g^{(c)}/\Gamma_g = 126.5\%$ of Pole~\#2 on $\text{RS}=5$ reflects through Eq.~(\ref{eq:R_first}) the energy variation of the rest of $\mathcal{S}_{cc}$ across its large width ($\Gamma_g=2.97\text{ MeV}$). The barrier penetrabilities of the channels that open at the $1p_{3/2}^{-1}$ threshold on $\text{RS}=5$ and the neighboring poles on the same sheet contribute to this variation. For the narrower Pole~\#2 on $\text{RS}=7$ at $36.64\text{ MeV}$ ($\Gamma_g=0.85\text{ MeV}$), $\sigma_g^{(c)}/\Gamma_g = 102.1\%$ is close to unity, as expected when the width is small compared with the energy scale over which the rest of $\mathcal{S}_{cc}$ varies.

\section{Summary and Conclusions}
\label{sec:summary}

In this work, we have investigated the microscopic structure and decay properties of nuclear giant resonances within the coordinate-space coupled-channel continuum RPA. 
By applying rank-1 Takagi factorization to the S-matrix residue at each resonance pole, we have obtained complex channel amplitudes that give the partial decay widths and relative branching ratios together with the channel phases. We have then factorized the complex transition strength of each pole into the single-particle scale, the participation rate, and the phase coherence, and constructed the dimensionless metric $C_g^{(\tau)2} N_{\text{tot}}\eta_g^{(\tau)}$, which equals unity in the single-particle limit, exceeds unity for coherent multi-channel enhancement, and falls below unity for single-channel dominance or destructive phase cancellation, without empirical width broadening or bound-state discretization.

The application of this methodology to $^{16}\text{O}$ yields several key physical insights across different multipole and isospin excitation channels:
\begin{enumerate}[(i)]
\item \emph{Isovector giant dipole resonance ($L=1, \tau=1$):} The prominent collective resonance on $\text{RS}=5$ at $21.72\text{ MeV}$ displays an enhanced collectivity factor of $3.492$ (with $N_{\text{tot}}\eta_g^{(1)} = 11.915$), providing strong evidence for its multi-channel collective character. The open-channel decay amplitudes show that all $10$ open channels participate concurrently, with proton emission ($71.1\%$) favored over neutron emission ($28.9\%$) and no single channel exceeding $28.6\%$.
\item \emph{Isoscalar quadrupole response and Fano antiresonance ($L=2, \tau=0$):} The attractive isoscalar residual interaction concentrates the collective ISGQR down to $16.77\text{ MeV}$ on $\text{RS}=3$ with a large collectivity index of $7.090$. 
As a consequence of this strength concentration, core-excited states at higher excitation energies ($27.03\text{ MeV}$ on $\text{RS}=5$) are left with virtually zero direct ground-state excitation amplitude. Their decay branching is overwhelmingly dominated ($82.32\%$) by deep inner-shell $1p_{3/2}^{-1}$ hole configurations (with $1p_{3/2} \to f_{7/2}$ transitions alone accounting for $79.99\%$). The algebraic relation between the effective Fano asymmetry parameter and the Jost residue phase, $q = -\cot\Theta_g \to 0$ ($\Theta_g \to \pm\pi/2$), provides a plausible microscopic picture of how destructive continuum interference suppresses the cross section at the resonance center, producing characteristic Fano-like antiresonance dips. The observed minimum of the isoscalar strength at $28.25\text{ MeV}$ is associated with the pole at $28.50\text{ MeV}$ ($q=+0.385$), and the pole at $30.08\text{ MeV}$ ($q\simeq-0.12$) is the closest to a pure antiresonance.
\item \emph{Isovector quadrupole response and fragmentation ($L=2, \tau=1$):} The prominent peak at $27.03\text{ MeV}$ on $\text{RS}=5$ exhibits a collectivity factor of $0.735$, which remains below the single-particle limit ($=1.0$). Its decay is heavily concentrated into a single proton transition ($p(1p_{3/2} \to f_{7/2})$ carrying $63.75\%$, with total proton decay reaching $81.1\%$), trapped by the steep $l=3$ centrifugal barrier ($l(l+1)=12$). This identifies the state as a quasi-single-particle continuum resonance rather than a collective mode. 
In contrast, collective IVGQR components according to the present diagnostic (collectivity factor $>1.0$) are pushed up and fragmented into higher-energy Riemann sheets ($30.08\text{ MeV}$ on $\text{RS}=5$ and $36.64\text{ MeV}$ on $\text{RS}=7$), with the latter displaying near-perfect isospin symmetry ($50.7\%$ neutron : $49.3\%$ proton). This microscopic picture offers a plausible physical scenario within the continuum RPA for understanding why experimental radiative capture studies observed broad, fragmented E2 strength distributions across $25\text{--}40\text{ MeV}$ rather than a single compact giant resonance.
\end{enumerate}

Finally, we emphasize that the continuum RPA used here incorporates multi-channel particle-hole configuration mixing, open-channel boundary conditions, and ground-state correlations through the backward amplitudes, but uses a phenomenological Woods-Saxon mean field and Skyrme residual interaction rather than a fully self-consistent framework, and is restricted to the $1p$-$1h$ configuration space without $2p$-$2h$ spreading widths $\Gamma^\downarrow$. The quantitative results, such as the pole energies, decay branching ratios, and collectivity indices, therefore depend on these model choices. 
Nevertheless, the rank-1 Takagi decomposition of the coupled-channel Jost-RPA S-matrix provides a systematic and physically transparent diagnostic framework for analyzing continuum resonance poles once a microscopic mean-field and residual interaction are specified.

Looking ahead, a natural extension of the present framework is its systematic application to medium and heavy closed-shell nuclei (such as $^{40}\text{Ca}$, $^{48}\text{Ca}$, and $^{208}\text{Pb}$), where giant collective modes are fully developed and large channel dimensions are required. 
Furthermore, an important future direction lies in investigating neutron-rich and drip-line nuclei, where low-energy continuum modes (such as the pygmy dipole resonance) emerge near weakly bound neutron thresholds. 
The present complex-pole decomposition and collectivity diagnostic metric provide an effective means to examine the microscopic nature of these structures—specifically, distinguishing collective vibrations of the neutron skin from unperturbed single-particle threshold resonances and clarifying their open-channel decay branching properties.

\appendix

\section{Jost function within the RPA formalism}
\label{app:jost_formalism}

We summarize the coordinate-space coupled-channel Jost function formalism. Within the continuum RPA framework \cite{Mizuyama2023_JRPA1, Mizuyama2024_Eigenphase, Mizuyama2025_JRPA2}, the regular matrix solution $\Phi^{(r)}(r, E)$ and irregular Jost matrix solutions $\Phi^{(\pm)}(r, E)$ of dimension $2N_{\text{tot}} \times 2N_{\text{tot}}$ satisfy the coupled radial Schrödinger-type system:
\begin{equation}
\left[ \mathbf{I} \frac{d^2}{dr^2} - \frac{\mathbf{L}^2}{r^2} + \mathcal{K}^2(E) - \frac{2m}{\hbar^2} \mathcal{V}(r) \right] \Phi^{(r,\pm)}(r, E) = 0,
\label{eq:coupled_radial}
\end{equation}
where $\mathbf{I}$ is the identity matrix, $\mathbf{L}^2$ is the diagonal matrix of the centrifugal barrier terms given by $\mathbf{L}^2 = \text{diag}\left(l_1(l_1+1), \dots, l_{N_{\text{tot}}}(l_{N_{\text{tot}}}+1), l_1(l_1+1), \dots \right)$, and where matrix $\mathcal{K}(E)$ is the diagonal momentum defined by:
\begin{equation}
\mathcal{K}(E) = \begin{pmatrix} \mathcal{K}_1(E) & \mathbf{0} \\ \mathbf{0} & \mathcal{K}_2(E) \end{pmatrix},
\label{eq:momentum_matrix}
\end{equation}
with the diagonal blocks:
\begin{align}
\mathcal{K}_1(E) &= \text{diag}\left( k_1^{(1)}, \dots, k_1^{(N_{\text{tot}})} \right), \\
\mathcal{K}_2(E) &= \text{diag}\left( k_2^{(1)}, \dots, k_2^{(N_{\text{tot}})} \right),
\end{align}
where the channel wave numbers are given by $k_1^{(i)}(E) = \sqrt{2m(\epsilon_{h_i} + E)/\hbar^2}$ and $k_2^{(i)}(E) = \sqrt{2m(\epsilon_{h_i} - E)/\hbar^2}$.
Here, the excitation energy $E$ resides in the complex plane. The channel space is doubled to $2N_{\text{tot}}$ to incorporate both forward and backward excitation amplitudes. The $2N_{\text{tot}} \times 2N_{\text{tot}}$ total potential matrix $\mathcal{V}(r)$ is structured as
\begin{equation}
  \mathcal{V}(r)
  = \begin{pmatrix}
    \mathcal{U}(r) + \mathcal{V}_{\text{res}}(r) & \mathcal{V}_{\text{res}}(r)
    \\
    \mathcal{V}_{\text{res}}(r) & \mathcal{U}(r) + \mathcal{V}_{\text{res}}(r)
  \end{pmatrix},
\label{eq:v_block}
\end{equation}
where $N_{\text{tot}} = N_n + N_p$ represents the total particle-hole channel dimension, and $\mathcal{U}(r) = \operatorname{diag}\left(U_{nl_1 j_1}(r),\dots, U_{pl_{N_p} j_{N_p}}(r)\right)$ contains the unperturbed single-particle mean-field potentials:
\begin{equation}
  U_{qlj}(r) = V_{\text{cent}}^{(q)}(r) + \bra \mathbf{l} \cdot \boldsymbol{\sigma} \ket V_{\text{so}}^{(q)}(r)
  + \delta_{qp}V_{\text{C}}(r),
\label{eq:v_lj}
\end{equation}
with the spin-orbit factor $\bra \mathbf{l} \cdot \boldsymbol{\sigma} \ket = j(j+1) - l(l+1) - 3/4$,
where $m$ is the nucleon mass, and $\boldsymbol{\sigma} = 2 \mathbf{s}$ is the Pauli spin operator.
The central potential $V_{\text{cent}}^{(q)}(r)$ and the spin-orbit potential $V_{\text{so}}^{(q)}(r)$
are parametrized with an isospin-dependent factor as
\begin{align}
  V_{\text{cent}}^{(q)}(r)
  &
  = V_0 \left( 1 - \kappa \bra q|\tau_z|q\ket \frac{N - Z}{A} \right)
  f_{\text{WS}}\left(\frac{r - R_0}{a_0}\right),
  \label{eq:vcent}
  \\
  V_{\text{so}}^{(q)}(r)
  &
  =- V_{\text{ls}} \left( 1 - \kappa \bra q|\tau_z|q\ket \frac{N - Z}{A} \right) \frac{1}{r a_{\text{ls}}}
  g_{\text{WS}}\left(\frac{r - R_{\text{ls}}}{a_{\text{ls}}}\right),
  \label{eq:vso}
\end{align}
where $V_0$ and $V_{\text{ls}}$ determine the strengths of the central and spin-orbit potentials, respectively, and $\kappa = 0.67$
is the parameter for the isospin asymmetry. The Woods-Saxon shape functions $f_{\text{WS}}(x)$ and $g_{\text{WS}}(x)$ are defined by
\begin{align}
  f_{\text{WS}}(x) &= \frac{1}{1 + \exp(x)},
  \label{eq:fws} \\
  g_{\text{WS}}(x) &= \frac{\exp(x)}{\left(1 + \exp(x)\right)^2},
  \label{eq:gws}
\end{align}
with the radii $R_0 = r_0 A^{1/3}$, $R_{\text{ls}} = r_{\text{ls}} A^{1/3}$ and diffuseness parameters $a_0$, $a_{\text{ls}}$.
The Coulomb potential $V_{\text{C}}(r)$ for protons, modeling a uniformly charged sphere of radius $R_{\text{C}} = r_{\text{c}} A^{1/3}$,
is expressed as
\begin{equation}
V_{\text{C}}(r) = 
\begin{cases}
\frac{Ze^2}{2 R_{\text{C}}} \left( 3 - \frac{r^2}{R_{\text{C}}^2} \right) & \text{for } r \le R_{\text{C}}, \\
\frac{Ze^2}{r} & \text{for } r > R_{\text{C}}.
\end{cases}
\label{eq:vcoul}
\end{equation}
and $\mathcal{V}_{\text{res}}(r)$ is the residual particle-hole interaction matrix defined by
\begin{equation}
  \mathcal{V}_{\text{res}}(r)
  =
  \sum_{qq'}
  \tilde{\varphi}_{q}(r)
  \frac{\kappa_{qq'}(r)}{r^2}
  \tilde{\varphi}^{\mathsf{T}}_{q'}(r)
\end{equation}
where $q, q' \in \{n, p\}$ label the isospin of the interacting nucleons.
The coordinate-dependent residual interaction $\kappa_{qq'}(r)$ is parametrized using the standard Skyrme force parameters
($t_0, x_0, t_3, x_3$) and the single-particle densities $\rho_n(r)$ and $\rho_p(r)$:
\begin{eqnarray}
  \kappa_{qq}(r)
  &=& \frac{1}{2} t_0 (1 - x_0)
  \nonumber\\
  &&
  + \frac{1}{12} t_3 \left[ (5 + x_3)\rho(r) - (2 + 4x_3)\rho_q(r) \right],
  \label{eq:vqq}
  \\
  \kappa_{qq'}(r)
  &=& t_0 \left( 1 + \frac{1}{2} x_0 \right) + \frac{1}{12} t_3 (5 + x_3)\rho(r),
  \label{eq:vqqbar}
\end{eqnarray}
where $\rho(r) = \rho_n(r) + \rho_p(r)$ is the total nucleon density, and $q'$ represents the isospin opposite to $q$.
$\tilde{\varphi}_{q}(r)$ is the $N_{\text{tot}}$-dimensional vector defined as
\begin{equation}
  \tilde{\varphi}_{n}(r)
  =
  \begin{pmatrix}
    \tilde{\varphi}_{n,1}(r) \\
    \vdots \\
    \tilde{\varphi}_{n,N_n}(r)\\
    0 \\
    \vdots\\
    0
  \end{pmatrix},
  \tilde{\varphi}_{p}(r)
  =
  \begin{pmatrix}
    0 \\
    \vdots\\
    0 \\
    \tilde{\varphi}_{p,1}(r) \\
    \vdots \\
    \tilde{\varphi}_{p,N_p}(r)\\
  \end{pmatrix},
  \label{holevec}
\end{equation}
with the component of the vector $\tilde{\varphi}_{q,\alpha}(r)$ is defined by
\begin{eqnarray}
  \tilde{\varphi}_{q,\alpha}(r)
  &=&
  \frac{\bra l_p^{(\alpha)}j_p^{(\alpha)}||Y_L||l_h^{(\alpha)}j_h^{(\alpha)}\ket}{\sqrt{2L+1}}
  \varphi_{q,\alpha}(r)
  \label{eq:resforce}
\end{eqnarray}
where $\varphi_{q,\alpha}(r)$ is a hole state wave function which is obtained by solving
\begin{equation}
  \left[ -\frac{\hbar^2}{2m} \frac{d^2}{dr^2} + \frac{\hbar^2 l(l+1)}{2m r^2} + U_{qlj}(r) \right]
  \varphi_{qi}(r) = e_{qi} \varphi_{qi}(r),
\label{eq:radial_se}
\end{equation}
$l_{p}$ and $j_{p}$ represent the angular momenta of the particle states coupled with
the hole states. 

The regular matrix solution $\Phi^{(r)}(r, E)$ satisfies the physical boundary condition at the origin:
\begin{equation}
\Phi^{(r)}_{\alpha\beta}(r, E) \sim \delta_{\alpha\beta} r j_{l_p^{(\alpha)}}(k_{\alpha}r) \quad (r \to 0)
\end{equation}
with $j_{l}(z)$ denoting the spherical Bessel function. The irregular Jost solutions $\Phi^{(\pm)}(r, E)$ are characterized by their asymptotic outgoing and incoming spherical Hankel wave conditions at large radial distances:
\begin{equation}
\Phi^{(\pm)}_{\alpha\beta}(r, E) \sim \delta_{\alpha\beta} r h_{l_\alpha}^{(1,2)}(k_\alpha r) \quad (r \to \infty),
\end{equation}
where $h_l^{(1)}(z)$ and $h_l^{(2)}(z)$ are the spherical Hankel functions representing outgoing
and incoming waves, respectively.

The Jost matrices $\mathcal{J}^{(\pm)}(E)$ serve as the connection matrices relating the regular and irregular solution bases:
\begin{eqnarray}
  \Phi^{(r)\mathsf{T}}(r, E)
  &=& \frac{1}{2}
  \left[ \mathcal{J}^{(+)}(E) \Phi^{(-)\mathsf{T}}(r, E)\right.
    \nonumber\\
    &&
    + \left.\mathcal{J}^{(-)}(E) \Phi^{(+)\mathsf{T}}(r, E) \right].
  \label{eq:jost_definition}
\end{eqnarray}
Introducing the generalized matrix Wronskian $\mathcal{W}\left(\Phi^{(1)}, \Phi^{(2)}\right)$ defined by
\begin{align}
  \mathcal{W}\left(\Phi^{(1)}, \Phi^{(2)}\right)
  &\equiv
  \frac{\hbar^2}{2m}
  \left[ \Phi^{(1)\mathsf{T}}(r) \left( \frac{\partial}{\partial r} \Phi^{(2)}(r) \right)\right.
    \nonumber\\
    &
    - \left.\left( \frac{\partial}{\partial r} \Phi^{(1)\mathsf{T}}(r) \right) \Phi^{(2)}(r) \right],
  \label{eq:wronskian_def}
\end{align}
which is coordinate-independent, the Jost function matrices are evaluated via the Wronskian relations:
\begin{equation}
  \mathcal{J}^{(\pm)}(E) = \mp i \frac{2m}{\hbar^2}
  \mathcal{W}\left(\Phi^{(r)}, \Phi^{(\pm)}\right) \mathcal{K}(E),
\label{eq:jost_wronskian}
\end{equation}
owing to the fundamental irregular solution Wronskian identity $\mathcal{W}\left(\Phi^{(-)}, \Phi^{(+)}\right) = i \frac{\hbar^2}{m} \mathcal{K}^{-1}$.

Because regular solutions vanish at the origin ($\Phi^{(r)}(r=0)=0$), their mutual Wronskian vanishes identically: $\mathcal{W}\left(\Phi^{(r)}, \Phi^{(r)}\right) = 0$. Combining this boundary property with Eq.~(\ref{eq:jost_definition}) directly yields the reciprocity relation for the Jost matrices:
\begin{equation}
  \mathcal{J}^{(+)}
  \mathcal{K}^{-1}
  \mathcal{J}^{(-)\mathsf{T}}
  =
  \mathcal{J}^{(-)}
  \mathcal{K}^{-1}
  \mathcal{J}^{(+)\mathsf{T}}.
  \label{eq:sympro-jost}
\end{equation}

\section{Jost-RPA Green's function, RPA response function and strength function}
\label{app:jost_green}

The coordinate-space Jost-RPA Green's function $\mathcal{G}(r,r')$ is defined as the resolvent kernel satisfying the inhomogeneous equation
\begin{align}
  &
  \left[
    \frac{\hbar^2}{2m}
    \mathcal{K}^2
    -
    \left\{
    -\frac{\hbar^2}{2m}
    \left(
    \del{^2}{r^2}
    -
    \frac{l(l+1)}{r^2}
    \right)
    \bvec{I}
    +
    \mathcal{V}(r)
    \right\}
    \right]
  \mathcal{G}(r,r')
  \nonumber\\
  &
  =
  \bvec{I}
  \delta(r-r')
  \label{defGreen}
\end{align}
and is constructed from the regular and irregular Jost solutions and the Jost matrix inversion as
\begin{align}
  \mathcal{G}(r,r')
  &=
  -i\frac{2m}{\hbar^2}
  \left[
    \theta(r-r')
    \Phi^{(+)}(r)
    \mathcal{K}
    \mathcal{J}^{(+)-1}
    \Phi^{(r)\mathsf{T}}(r')
    \right.
    \nonumber\\
    &
    +
    \left.
    \theta(r'-r)
    \Phi^{(r)}(r)
    \left(\mathcal{J}^{(+)-1}\right)^{\mathsf{T}}
    \mathcal{K}
    \Phi^{(+)\mathsf{T}}(r')
    \right].
  \label{Green}
\end{align}
The physical scattering wave function $\Psi^{(+)}(r;E)$ is defined through the outgoing boundary condition as
\begin{align}
  \Psi^{(+)\mathsf{T}}(r;E)
  &\equiv
  \mathcal{J}^{(+)-1}(E)
  \Phi^{(r)\mathsf{T}}(r;E)
  \label{defpsi0}
  \\
  &=
  \frac{1}{2}
  \left[
    \Phi^{(-)\mathsf{T}}(r;E)
    +
    S(E)
    \Phi^{(+)\mathsf{T}}(r;E)
    \right]
  \label{defpsi}
\end{align}
where the unnormalized $S$-matrix is $S(E) \equiv \left(\mathcal{J}^{(+)}\right)^{-1} \mathcal{J}^{(-)}$. Consequently, the resonance poles $E_g$ are determined by the discrete secular condition:
\begin{equation}
  \det\mathcal{J}^{(+)}(E_g)=0.
\end{equation}

By defining the flux-adjusted symmetric scattering matrix $\mathcal{S}(E)$ through kinematic momentum scaling:
\begin{equation}
  \mathcal{S}(E)
  \equiv
  \mathcal{K}^{\frac{1}{2}}
  S(E)
  \mathcal{K}^{-\frac{1}{2}}
  =
  \mathcal{K}^{\frac{1}{2}}
  \left(\mathcal{J}^{(+)}\right)^{-1}
  \mathcal{J}^{(-)}
  \mathcal{K}^{-\frac{1}{2}}
  \label{eq:Smat-F}
\end{equation}
the resulting scattering matrix is symmetric: $\mathcal{S}^{\mathsf{T}}(E) = \mathcal{S}(E)$. Introducing the flux-adjusted radial wave solutions
\begin{align}
  \Phi^{(r)\mathsf{T}}_F(r;E)
  &\equiv
  \mathcal{K}^{\frac{1}{2}}
  \Phi^{(r)\mathsf{T}}(r;E)
  \\
  \Phi^{(\pm)\mathsf{T}}_F(r;E)
  &\equiv
  \mathcal{K}^{\frac{1}{2}}
  \Phi^{(\pm)\mathsf{T}}(r;E),
\end{align}
the flux-adjusted scattering state becomes
\begin{align}
  \Psi^{(+)\mathsf{T}}_F(r;E)
  &\equiv
  \mathcal{K}^{\frac{1}{2}}
  \Psi^{(+)\mathsf{T}}(r;E)
  \label{defpsi0-F}
  \\
  &=
  \frac{1}{2}
  \left[
    \Phi^{(-)\mathsf{T}}_F(r;E)
    +
    \mathcal{S}(E)
    \Phi^{(+)\mathsf{T}}_F(r;E)
    \right],
  \label{defpsi-F}
\end{align}
and the Green's function admits the compact symmetric representation:
\begin{align}
  \mathcal{G}(r,r')
  &=
  -i\frac{2m}{\hbar^2}
  \left[
    \theta(r-r')
    \Phi^{(+)}_F(r)
    \Psi^{(+)\mathsf{T}}_F(r')
    \right.
    \nonumber\\
    &
    +
    \left.
    \theta(r'-r)
    \Psi^{(+)}_F(r)
    \Phi^{(+)\mathsf{T}}_F(r')
    \right].
  \label{Green-F1}
  \\
  &=
  -i\frac{m}{\hbar^2}
  \left[
    \theta(r-r')
    \Phi^{(+)}_F(r)
    \Phi^{(-)\mathsf{T}}_F(r')
    \right.
    \nonumber\\
    +
    &
    \theta(r'-r)
    \Phi^{(-)}_F(r)
    \Phi^{(+)\mathsf{T}}_F(r')
    \nonumber\\
    &+
    \Phi^{(+)}_F(r)
    \mathcal{S}(E)
    \Phi^{(+)\mathsf{T}}_F(r')
    \left.
    \right].
  \label{Green-F2}
\end{align}
Using the $2N_{\text{tot}}$-dimensional hole-state vector $\xi_q(r)$ built from $\tilde{\varphi}_{q}(r)$ [Eq.~(\ref{holevec})]:
\begin{equation}
  \xi_q(r)
  \equiv
  \begin{pmatrix}
    \tilde{\varphi}_{q}(r) \\
    \tilde{\varphi}_{q}(r)
  \end{pmatrix},
\end{equation}
the coordinate-space RPA response function $R_{qq'}(r,r';E)$ is given by
\begin{align}
  R_{qq'}(r,r')
  &=
  \xi_q^{\mathsf{T}}(r)
  \mathcal{G}(r,r')
  \xi_{q'}(r').
  \label{eq:Rqq}
\end{align}
The continuum RPA strength function $S_F^{(\tau)}(E)$ is obtained from the imaginary part of the integrated multipole response $R_F(E)$:
\begin{equation}
  S_F^{(\tau)}(E)
  \equiv
  -\operatorname{Im}R_F(E)
  \label{eq:strength}
\end{equation}
where
\begin{equation}
  R_F(E)
  \equiv
  \frac{1}{\pi}
  \sum_{qq'}
  \int\int drdr'
  f_{q}^{(\tau)}(r)
  R_{qq'}(r,r';E)
  f_{q'}^{(\tau)}(r')
  \label{eq:defRF}
\end{equation}
and $f_{q}^{(\tau)}(r)$ is the external multipole transition operator, defined by
\begin{align}
  f_{q}^{(\tau)}(r)
  &=
  \left\{
  \begin{array}{cc}
    r^{L} & \mbox{$\tau=0$ (Isoscalar)} \\
    \tau_z r^{L} & \mbox{$\tau=1$ (Isovector)} 
  \end{array}
  \right.
  \label{eq:f_general}
\end{align}
when $L=1$
\begin{align}
  f_{q}^{(\tau)}(r)
  &=
  \left\{
  \begin{array}{cc}
    r^{3}-\frac{5}{3}\bra r^2\ket r & \mbox{$\tau=0$ (Isoscalar)} \\
    \delta_{qn}\frac{Z}{A}r-\delta_{qp}\frac{N}{A}r & \mbox{$\tau=1$ (Isovector)} 
  \end{array}
  \right..
  \label{eq:f_dipole}
\end{align}

\begin{acknowledgments}
This work was supported by the Computer System for Nuclear Physics at the Research Center for Nuclear Physics (RCNP), Osaka University.
\end{acknowledgments}

\end{document}